\documentclass[]{jfm}
\usepackage{graphicx}
\usepackage{epstopdf, epsfig}
\usepackage{newtxtext}
\usepackage[slantedGreek]{newtxmath}
\usepackage{natbib}
\usepackage{hyperref}
\usepackage{amsmath}
\usepackage{subfigure}
\usepackage{upgreek}
\usepackage[export]{adjustbox}
\usepackage[table,usenames,dvipsnames]{xcolor}
\usepackage[normalem]{ulem}
\usepackage[ruled,vlined]{algorithm2e}

\def\lline{\vrule width14pt height2.5pt depth -2pt}

\def\m1line{\vrule width3pt height2.5pt depth -2pt}

\def\thickline{\vrule width14pt height3.5pt depth -2pt}
\def\bdot{\raise.2em\hbox to .15em{.}}

\def\graythickline{\textcolor{gray}{\thickline}}

\def\dashed{\m1line\hskip3.5pt\m1line\hskip3.5pt\m1line\thinspace}

\def\dotted{\bdot\ \bdot\ \bdot\ \bdot\thinspace}

\def\outline#1{\textcolor{Gray}{}}
\long\def\comment#1{}

\shorttitle{The origin of vorticity in viscous flows}
\shortauthor{T. Xiang, G. Eyink \& T. A. Zaki} 

\title{The origin of vorticity in viscous incompressible flows}
\author{Tianrui Xiang\aff{1}, Gregory L. Eyink\aff{1,2} \and Tamer A. Zaki\aff{1,2}\corresp{\email{t.zaki@jhu.edu}}}
\affiliation{\scriptsize
\aff{1}Department of Mechanical Engineering, Johns Hopkins University, Baltimore, MD 21218, USA.  
\aff{2}Department of Applied Mathematics and Statistics, Johns Hopkins University, Baltimore, MD 21218, USA
}

\begin{document}
	
\maketitle
	
\begin{abstract}

In inviscid, incompressible flows, the evolution of vorticity is exactly equivalent to that of an infinitesimal material line-element, and hence vorticity can be traced forward or backward in time in a Lagrangian fashion. 
This elegant and powerful description is not possible in viscous flows due to the action of diffusion. 
Instead, a stochastic Lagrangian interpretation is required and was recently introduced, where the origin of vorticity at a point is traced back in time as an expectation over the contribution from stochastic trajectories. 
We herein introduce for the first time an Eulerian, adjoint-based approach to quantify the back-in-time origin of vorticity in viscous, incompressible flows. 
The adjoint variable encodes the advection, tilting and stretching of the earlier-in-time vorticity that ultimately leads to the target value.  Precisely, the adjoint vorticity is the volume-density of the mean Lagrangian deformation of the earlier vorticity.  The formulation can also account for the injection of vorticity into the domain at solid boundaries.  We demonstrate the mathematical equivalence of  the adjoint approach and the stochastic Lagrangian approach.  We then provide an example from turbulent channel flow, where we analyze the origin of high-stress events and relate them to Lighthill's mechanism of stretching of near-wall vorticity.               
\end{abstract}

\begin{keywords}
	Vorticity, Turbulence
\end{keywords}

\section{Introduction}
\label{sec:intro}

Vorticity and vortex lines have often been advocated  as key to the very difficult problem of fluid turbulence, 
by scientists such as G. I. Taylor \citep{taylor1932transport,taylor1937mechanism,taylor1938production},
M. J. Lighthill 
\citep{Lighthill} and 
G. L. Brown \& A. Roshko 
\citep{brown1974density,brown2012turbulent}. Quoting directly from the latter authors:   
\begin{quote} 
    \noindent
    \emph{``An understanding of the mechanics is most likely to be obtained from the vorticity. The subject stands at the beginning of a new era in which both LES and DNS calculations can provide details of the vorticity field and the fluxes of vorticity (vortex force).''}  \citep{brown2012turbulent}.     
\end{quote}
Much of the power and appeal of vorticity arises from its deep Lagrangian properties for ideal fluids, known since the $19^{\textrm{th}}$ century from the classical works by \cite{cauchy1815theorie}, \cite{helmholtz1858uber}, \cite{weber1868ueber}, and \cite{kelvin1868vi}. Unfortunately, the ideal Lagrangian vorticity invariants of Cauchy and Kelvin-Helmholtz are not conserved for real-world viscous flows, even in the limit of very high Reynolds number where the viscous effects might na\"ively be assumed to negligible. Several experimental and numerical studies have verified that the material properties of vortex lines for ideal flow are not observed to hold in such turbulent flows \citep{luthi2005lagrangian,guala2005evolution,guala2006stretching,chen2006kelvin,johnson2016large}.  
It is of course easy to incorporate viscosity into the Eulerian description of vorticity by the Helmholtz equation. However, the direct connection to the ideal vorticity dynamics is then lost and intuitive Lagrangian arguments for turbulent flow such as those made by \citet{taylor1937mechanism,taylor1938production}, and by \citet{Lighthill} appear then baseless and doubtful.

\begin{figure}
    \centering
    \includegraphics[width=\textwidth]{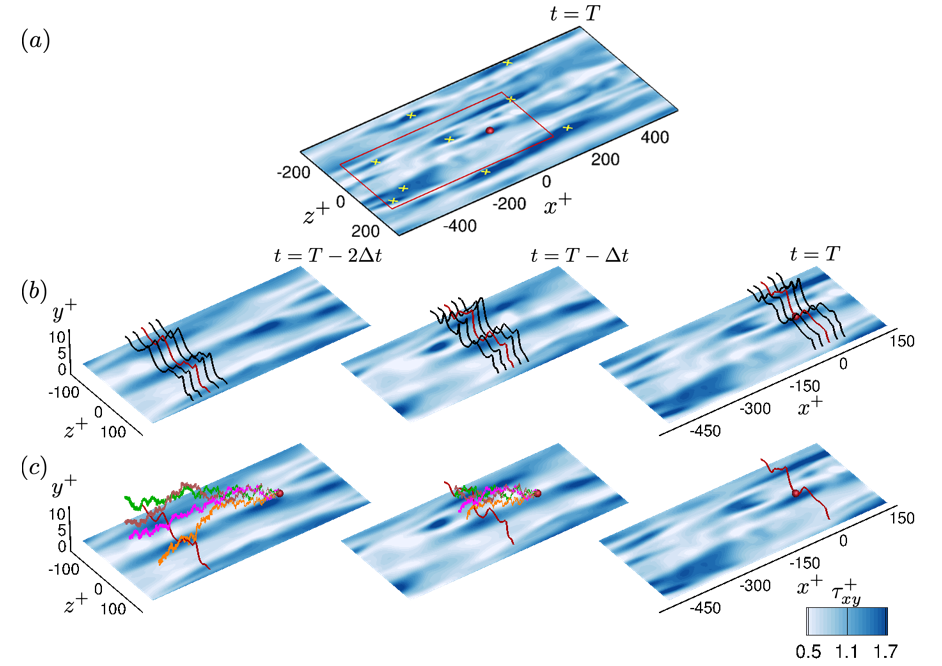}
    \caption{Contours of streamwise wall shear stress in turbulent channel flow at $Re_{\tau}=180$.  $(a)$ Stress at final time $t=T$, with symbols marking the local maxima. $(b)$ Schematic of the evolution of vortex lines in the inviscid viewpoint at three instances $t = \{T-2\Delta t,~T-\Delta t,~T\}$.  Five vortex lines are plotted in each frame.   $(c)$ Stochastic trajectories in backward time for the viscous flow, released at $t=T$ from a point along the vortex line above a stress maximum.}
    \label{fig:schematic}
\end{figure}

A fundamental advance was made, in our opinion, by \citet{ConstantinIyer08} and \citet{Constantin2011}, who discovered that the remarkable Lagrangian properties of vorticity for ideal Euler flows carry over to viscous Navier-Stokes in a stochastic formulation. 
In this approach, the equations for deterministic Lagrangian particle trajectories are perturbed by a Brownian noise which represents the viscous diffusion of vorticity, and averaging over the realizations of this process then recovers the vorticity of the viscous Navier-Stokes solution.  This stochastic approach is widely used to represent diffusion in applied mathematics \citep{oksendal2013stochastic} and also in engineering modelling \citep{sawford2001turbulent} and theoretical physics analysis
\citep{bernard1998slow} of passive scalar advection. 
The original paper by \citet{ConstantinIyer08} for flows in a periodic domain considered forward-in-time stochastic trajectories and averaged all vorticity vectors arriving, stretched and rotated, to the same final point. 
The second work by \citet{Constantin2011} considered wall-bounded flows and used an equivalent formulation by backward-in-time stochastic trajectories which all emanate from the target point, with vorticity vectors then transported forward along these paths and averaged.
The conceptual difference between the inviscid and viscous viewpoints is illustrated in figure \ref{fig:schematic}.  Vortex lines are plotted at three time instances in the vicinity of the wall in turbulent channel flow.  In the top panels, the na\"ive inviscid interpretation regards the red line as a material line.  The correct, stochastic Lagrangian interpretation is in the bottom panels.  The vorticity identified by the red point at the terminal time is traced back using the stochastic trajectories, four of which are shown. The earlier vorticity vectors at the tips of these trajectories are then transported forward, tilted and stretched, and finally averaged to make up the target value.  
This fact corresponds to validity of a stochastic Kelvin Theorem and a stochastic Cauchy invariant backward in time for viscous Navier-Stokes solutions \citep{eyink2010stochastic}. 
The stochastic trajectories also illustrate that, in a viscous flow, the vorticity at the final target point has contributions from a large spatial domain relative to the view from ideal flows. 

A Monte Carlo numerical scheme to identify the origin of vorticity exploiting such backward stochastic trajectories was developed by \cite{Eyink2020_theory,Eyink2020_channel}. They examined the precursors of a low- and a high-vorticity event in turbulent channel flow at $Re_\tau=1000$, located within five viscous units above the wall.  The use of Dirichlet boundary conditions on the vorticity implied that stochastic Lagrangian trajectories stopped once they first reached the boundary. 
As such, these boundary conditions, unfortunately, do not permit the investigation of the origin of the wall-vorticity itself, and hence the wall stress. 
The following paper by \cite{wang_eyink_zaki_2022} introduced the Neumann boundary conditions on the vorticity, by using stochastic Lagrangian trajectories
reflected at the wall to sample the boundary vorticity source of \cite{Lighthill}. 
Furthermore,  \cite{wang_eyink_zaki_2022} exploited this new stochastic representation numerically to investigate 
the origin of the high wall-stress that is observed in transition to turbulence. 
They were able to verify a conjecture by \cite{Lighthill} that strong concentration of vorticity at the boundary must result from spanwise stretching of spanwise vortex lines.

Despite the scientific success of the stochastic Lagrangian representation, the Monte Carlo method 
suffers from a slow convergence rate, 
which is a serious numerical limitation.  Errors vanish only $\propto \sigma_{\Omega}/\sqrt{N}$ where $N$ is the number of sample realizations and $\sigma_\Omega$ is the standard deviation of the stochastic Cauchy invariant (whose mean is conserved). 
This is the same slow convergence which plagues random walk approaches for introducing viscosity in direct vortex methods to solve Navier-Stokes \citep{mimeau2021review}. 
The problem is made much worse by the Lagrangian chaos observed in turbulent wall-bounded flow \citep{johnson2016large}, because the exponential growth of vorticity in individual realizations leads to an exponential growth of $\sigma_\Omega$ backward in time. This error growth thus requires exponentially large sample sizes $N$ to obtain converged results.

The Lagrangian interpretation of the origin of vorticity suggests than an Eulerian formulation may be possible, although none has been derived in the literature.  In this work, we will pursue such Eulerian, back-in-time description of the origin of vorticity using adjoint techniques. While adjoint methods are commonly adopted to compute gradients, for example in optimization, flow control, and data assimilation \cite[e.g.][]{giles1997adjoint,Bewley2001,wang2021state,zaki2024analreview}, here will be seeking to evaluate the full contributions of stretching and tilting \citep{Orszag_Patera_1983} of earlier vorticity to the target event.  
Such adjoint representations can be constructed in many different ways and it is unclear why any one should be preferred over others. 
We  must therefore ensure that the derived equations are mathematically equivalent to the Lagrangian representation.  
We will also provide a clear physical interpretation of the adjoint representation, and relate the mathematical expression precisely to the fluid dynamics.  
A key advantage of an Eulerian approach will be that it requires only the solution of deterministic partial differential equations, which can be accomplished by standard numerical discretization methods, and avoids completely the slow convergence problems of Monte Carlo methods.  

The derivation of the Eulerian back-in-time vorticity equation in introduced in \S\ref{sec:adjoint}, and its equivalence to the stochastic Lagrangian representation is shown mathematically in \S\ref{sec:stochastic}.  We then provide a concrete application of the method in \S\ref{sec:channel}, in turbulent channel flow where we examine the origin of the vorticity at five viscous units above maxima of the streamwise wall shear stress as well as the origin of the stress maxima themselves.  Our analysis quantifies the contributions of tilting and stretching of the earlier vorticity and of the wall-fluxes. 
Our main  physical conclusions are twofold: that strong near-wall vorticity arises primarily from spanwise stretching of interior spanwise vorticity and to a lesser extent from the vorticity flux at the boundary. 
The inefficacy of the latter mechanism is explained using the concept of phase speed of physical fields \citep{Sreenivasan_1988,kim1993propagation}, specifically of the wall-vorticity flux relative to the adjoint variable, and their relative dephasing which will be explained in detail.
These results, as summarized in \S\ref{sec:conclusion}, illustrate the fundamental new insights on vorticity dynamics that can be obtained from our novel adjoint scheme.


\section{The origin of vorticity in backward time}

Our goal is to derive an adjoint equation that relates the terminal vorticity vector $\boldsymbol{\omega}(\boldsymbol{x}_f,T)$ at a target point in space and time to the vorticity field $\boldsymbol{\omega}(\boldsymbol{x},s)$ at an earlier time $s<T$. We shall do this in subsection \ref{sec:adjoint} and then 
we explain the equivalence of this adjoint representation to the stochastic Lagrangian representation in subsection \ref{sec:stochastic}. 

\subsection{The adjoint representation of vorticity}
\label{sec:adjoint}

We start from the Helmholtz equation that governs the forward evolution of vorticity,
\begin{eqnarray}
    \frac{\partial \boldsymbol{\omega}}{\partial t} + \boldsymbol{u}\cdot\boldsymbol{\nabla}\boldsymbol{\omega}-\boldsymbol{\omega}\cdot\boldsymbol{\nabla} \boldsymbol{u}-\nu\boldsymbol{\nabla}^2 \boldsymbol{\omega}=
        \boldsymbol{0}, 
\label{eq:vort_eq}
\end{eqnarray}
where $\boldsymbol{\omega} = \boldsymbol{\nabla} \times \boldsymbol{u}$ is the vorticity and $\boldsymbol{u}$ is the velocity, and $\nu$ is the fluid kinematic viscosity.  
This parabolic partial differential equation is not time-reversible due to the viscous term. As such, it cannot be solved backward in time to obtain vorticity at an earlier time from its final value at time $T$. 
In addition, a standard approach to derive an adjoint to the forward system would start by introducing an additional evolution equation for the velocity.   Rather than adopt this standard approach, we introduce a key assumption and derive an adjoint equation that will be shown to be equivalent to the stochastic Cauchy invariant.  Specifically, we will freeze the forward trajectory of the velocity $\boldsymbol{u}(\boldsymbol{x},t)$, or assume its knowledge.  This step is different from the standard adjoint derivation where variations are considered for all state variables 
\citep{giles1997adjoint,luchini2014adjoint,luchini2024introduction}.  Here, we are not concerned with variations in the vorticity due to change in the flow trajectory in state space.  Instead, we are concerned with the ``origin of vorticity" backward in time, along a frozen trajectory in state space.  This assumption was inspired by, and is fully consistent with, the stochastic formulation as will be shown in the next section.

With the above assumption, the standard procedure for deriving forward-adjoint duality yields the relation,  
	\begin{eqnarray}
	\label{eq:byparts}
        \begin{aligned}
         \left(\boldsymbol{\Omega}, \frac{\partial \boldsymbol{\omega}}{\partial t}+\boldsymbol{u} \cdot \boldsymbol{\nabla} \boldsymbol{\omega}-\boldsymbol{\omega} \cdot \boldsymbol{\nabla} \boldsymbol{u}-\nu \boldsymbol{\nabla}^2 \boldsymbol{\omega}\right)  =\left(\boldsymbol{\omega}, \frac{\partial \boldsymbol{\Omega}}{\partial \tau}-\boldsymbol{u} \cdot \boldsymbol{\nabla} \boldsymbol{\Omega}-\boldsymbol{\nabla} \boldsymbol{u} \cdot \boldsymbol{\Omega}-\nu \boldsymbol{\nabla}^2 \boldsymbol{\Omega}\right)  & \\
         +\int_D \boldsymbol{\Omega}\left(\boldsymbol{x},T\right) \cdot \boldsymbol{\omega}\left(\boldsymbol{x},T\right) dV -\int_D \boldsymbol{\Omega}\left(\boldsymbol{x},s\right) \cdot \boldsymbol{\omega}\left(\boldsymbol{x},s\right) dV ~~~ & \\
         + \int_s^T \oint_{\partial D} 
         \left[\left(\boldsymbol{\Omega} \cdot \boldsymbol{\omega}\right) \boldsymbol{u}\cdot \boldsymbol{n} 
         - 
         \nu  \left( \left( \boldsymbol{n}\cdot \boldsymbol{\nabla} \boldsymbol{\omega} \right) \cdot \boldsymbol{\Omega} 
	             - \left( \boldsymbol{n}\cdot \boldsymbol{\nabla} \boldsymbol{\Omega}\right) \cdot \boldsymbol{\omega}  \right)\right] \, dS \, dt.  ~~~ &
        \end{aligned}
	\end{eqnarray}
In the above expression, the adjoint variable is the vector $\boldsymbol{\Omega}$, the space-time inner product is $\left( \boldsymbol{f}, \boldsymbol{g}\right) = \int_s^T \int_D \boldsymbol{f}\cdot \boldsymbol{g} \, dV \, dt$, $dV$ is the volume element in the flow domain $D,$ $dS$ is the area element on the boundary surface $\partial D,$ $\boldsymbol{n}$ is the outward pointing surface normal, and $\tau = T-t$ is the reverse time.
The derivation can be considered as a variational calculation, in which $\boldsymbol{\Omega}$ represents a field of Lagrange multipliers to enforce the Helmholtz equation. Variation over  $\boldsymbol{\omega}$ then yields the the evolution equation for the adjoint vorticity,  
\begin{equation}
        \frac{\partial \boldsymbol{\Omega}}{\partial \tau} - \boldsymbol{u}\cdot\boldsymbol{\nabla}\boldsymbol{\Omega} - \boldsymbol{\nabla} \boldsymbol{u} \cdot \boldsymbol{\Omega} -\nu\boldsymbol{\nabla}^2 \boldsymbol{\Omega} = \boldsymbol{0}. 
\label{eq:advort_eq}
\end{equation}
The remaining terms relate the vorticity field at the final time $T$ to the 
initial vorticity field at time $s$ and the vorticity boundary conditions  
over the interval $\left[s,T\right]$: 
\begin{eqnarray}
\label{eq:duality}
\begin{aligned}
	\int_D \boldsymbol{\Omega}\left(\boldsymbol{x},T\right) \cdot \boldsymbol{\omega}\left(\boldsymbol{x},T\right) dV 
	 = & \int_D \boldsymbol{\Omega}\left(\boldsymbol{x},s\right) \cdot \boldsymbol{\omega}\left(\boldsymbol{x},s\right) dV ~~~  \\
         & + \int_s^T \oint_{\partial D} 
         \left[
         \nu  \left( \left( \boldsymbol{n}\cdot \boldsymbol{\nabla} \boldsymbol{\omega} \right)  \cdot \boldsymbol{\Omega} 
         	     - \left( \boldsymbol{n}\cdot \boldsymbol{\nabla} \boldsymbol{\Omega}\right) \cdot \boldsymbol{\omega} \right)\right]~dS~dt,
\end{aligned}
 \end{eqnarray}
where we have discarded the term $\int_s^T \oint_{\partial D} \left(\boldsymbol{\Omega} \cdot \boldsymbol{\omega}\right) \boldsymbol{u}\cdot \boldsymbol{n}~dS dt$ because it vanishes  at periodic boundaries, at no-penetration walls, and at the far field if the velocity decays. 
To obtain a useful representation of vorticity from (\ref{eq:duality}), we must impose additional initial and boundary conditions for $\boldsymbol{\Omega}$.

We now denote as $\boldsymbol{\Omega}^k$ the particular adjoint field which enables us to trace back from the $k^{\textrm{th}}$ component of the vorticity vector at a point in space $\boldsymbol{x}_f$ and time $T$ to its origin.  In other words, $\boldsymbol{\Omega}^k$ will capture how earlier vorticity $\boldsymbol{\omega}\left(\boldsymbol{x},s\right)$ was tilted, stretched and diffused to generate $\omega_{k}\left(\boldsymbol{x}_{f},T\right)$.  This adjoint field is defined at $t=T$ according to,
\begin{equation}
	\boldsymbol{\Omega}^k\left(\boldsymbol{x},T\right) = \boldsymbol{e}_k \delta\left(\boldsymbol{x}-\boldsymbol{x}_f\right), 
\end{equation}
which, substituted in equation (\ref{eq:duality}), yields the relation between the terminal and initial vorticity, 
\begin{eqnarray}
\label{eq:dualityk}
\begin{aligned}
	\omega_k\left(\boldsymbol{x}_f,T \right)  =& \int_D \boldsymbol{\Omega}^k\left(\boldsymbol{x},s\right) \cdot \boldsymbol{\omega}\left(\boldsymbol{x},s\right) dV ~~~  \\
         & + \int_s^T \oint_{\partial D} 
         \nu  \left(   \left( \boldsymbol{n}\cdot \boldsymbol{\nabla} \boldsymbol{\omega} \right)  \cdot \boldsymbol{\Omega}^k 
         	     - \left( \boldsymbol{n}\cdot \boldsymbol{\nabla} \boldsymbol{\Omega}^k\right) \cdot \boldsymbol{\omega} \right) ~dS dt,
\end{aligned}
\end{eqnarray}
Hence $\boldsymbol{\Omega}^k,$ $k=1,2,3$ are Eulerian vector fields that map the corresponding $k^{\textrm{th}}$ component of vorticity to its origin in a viscous flow.   

In presence of solid boundaries, a choice of homogeneous Dirichlet 
$\boldsymbol{\Omega}^k=\boldsymbol{0}$ 
or homogeneous Neumann 
$\boldsymbol{n}\cdot \boldsymbol{\nabla} \boldsymbol{\Omega}^k=\boldsymbol{0}$ boundary conditions can be adopted for the adjoint vorticity.  
The Dirichlet condition leaves a surface integral  
$- \int_s^T \oint_{\partial D} \nu \left( \boldsymbol{n}\cdot \boldsymbol{\nabla} \boldsymbol{\Omega}^k\right) \cdot \boldsymbol{\omega}~dS~dt$
where the wall-normal flux of the adjoint samples the wall vorticity over time.  
The formula \eqref{eq:dualityk} then yields the solution at time $t=T$ of the Helmholtz equation 
\eqref{eq:vort_eq}, with initial data at time $t=s$ and Dirichlet boundary conditions 
for vorticity over the interval $[s,T].$ On the other hand, the zero Neumann condition leaves a surface integral  
$ \int_s^T \oint_{\partial D} \nu\left( \boldsymbol{n}\cdot \boldsymbol{\nabla} \boldsymbol{\omega}\right) \cdot \boldsymbol{\Omega}^k ~dS~dt$ 
where the adjoint field samples the diffusive influx of vorticity at the wall  $\nu\boldsymbol{n}\cdot \boldsymbol{\nabla} \boldsymbol{\omega}$ over time. 
In that case, the formula \eqref{eq:dualityk} then yields the solution at time $t=T$ of the Helmholtz equation \eqref{eq:vort_eq}, with initial data at time $t=s$ and Neumann boundary conditions for vorticity over the interval $[s,T].$ 
Note that either boundary condition for the adjoint variable can be adopted for tracing back the origin of the vorticity within the bulk.  However, only the Neumann version is compatible with the initial condition $\boldsymbol{\Omega}^k\left(\boldsymbol{x},T\right) = \boldsymbol{e}_k \delta\left(\boldsymbol{x}-\boldsymbol{x}_{f}\right)$ for a point on the wall, and hence only the Neumann boundary condition can be adopted for tracing back the origin of the wall vorticity, or equivalently of the wall stress.  

The representation \eqref{eq:dualityk} is not unique. 
Other duality relations can be derived starting from different forms of the vorticity equation, for example by replacing the viscous term by $\nu\boldsymbol{\nabla}\times\boldsymbol{\nabla}\times\boldsymbol{\omega}$ and repeating the derivation. 
As we show in the next section \S\ref{sec:stochastic}, however, the representation \eqref{eq:dualityk} is distinguished by the fact that it is equivalent to the stochastic Lagrangian representation of \cite{ConstantinIyer08}.  
To facilitate the comparison, we gather the three adjoint 
fields for $k=1,2,3,$ interpreted as column vectors, to form the rows of a matrix ${\boldsymbol \Upomega}=
[{\boldsymbol \Omega}^1,{\boldsymbol \Omega}^2,{\boldsymbol \Omega}^3]^\top$, which satisfies,
\begin{eqnarray}
        \frac{\partial \boldsymbol{\Upomega}}{\partial \tau} - \boldsymbol{u}\cdot\boldsymbol{\nabla}\boldsymbol{\Upomega} - 
        \boldsymbol{\Upomega}(\boldsymbol{\nabla} \boldsymbol{u})^\top - \nu\boldsymbol{\nabla}^2\boldsymbol{\Upomega} = 0. 
    \label{eq:Omega_eq} 
\end{eqnarray}
Rewriting the duality relation \eqref{eq:dualityk},
we see that ${\boldsymbol \Upomega}$ acts on the initial vorticity vector and boundary sources to give the vorticity vector at the target position and time by the formula,
\begin{equation}
\label{eq:operator}
	\boldsymbol{\omega}\left(\boldsymbol{x}_f,T \right)  = \int_D {\boldsymbol \Upomega}\left(\boldsymbol{x},s\right) \boldsymbol{\omega}\left(\boldsymbol{x},s\right) dV 
         + \int_s^T \oint_{\partial D} 
         \nu  \left( {\boldsymbol \Upomega} \left( \boldsymbol{n}\cdot \boldsymbol{\nabla} \boldsymbol{\omega} \right)    
         	     - \left( \boldsymbol{n}\cdot \boldsymbol{\nabla} {\boldsymbol \Upomega}\right) \boldsymbol{\omega} \right)~dS~dt. 
\end{equation}
The physical interpretation of the Eulerian matrix ${\boldsymbol \Upomega}$ is most important.  
The field $\boldsymbol{\Upomega}(\boldsymbol{x},t)$ is the volume density of mean deformation experienced by vorticity from space-time point $(\boldsymbol{x},t)$ to the target point $(\boldsymbol{x}_f,T)$, as will emerge from the proof of equivalence in the following section.
In particular, for a smooth Euler solution obtained in the inviscid limit, the matrix ${\boldsymbol \Upomega}$ coincides with the deformation tensor $\boldsymbol{D}$ of standard continuum mechanics. Another instructive connection in the inviscid limit is to the evolution of an infinitesimal fluid element with volume $\delta V = \delta \boldsymbol{l} \cdot \delta \boldsymbol{A}$, where $\delta \boldsymbol{l}$ is the line element and $\delta \boldsymbol{A}$ is the area vector element.  The forward vorticity vector satisfies the same evolution equation as $\delta \boldsymbol{l}$, which is the foundation for our intuition regarding vorticity tilting and stretching.  Each row of ${\boldsymbol \Upomega}$ is an adjoint-vorticity vector $\boldsymbol{\Omega}$, which satisfies the same evolution equation as $\delta \boldsymbol{A}$ (see eq.\,(3.1.5) in 
\citet{Batchelor_2000}, which is identical to equation \eqref{eq:advort_eq} with $\tau=-t$ and $\nu=0$). Therefore, terms such as $\left(-\boldsymbol{\nabla} \boldsymbol{u} \cdot \boldsymbol{\Omega}\right)$ have an intuitive physical interpretation as stretching and tilting of the area vector element, i.e.\,enlarging and rotating the area.

In summary, ${\boldsymbol \Upomega}$ represents the action of advection, stretching, tilting and and also viscous diffusion on the initial vorticity and boundary source, which transforms them to produce the target vorticity vector.  As we will see by aid of an example from channel flow in \S\ref{sec:channel}, this interpretation permits one to understand intuitively the adjoint field and to connect its behaviour with the Lagrangian dynamics of the flow vorticity. 

In section \ref{sec:stochastic}, we very succinctly review the stochastic Lagrangian representation and explain its connection with the adjoint-vorticity equations 
\eqref{eq:Omega_eq}. The proof requires basic knowledge of stochastic calculus, and can be skipped without loss of continuity. The most important consequence is a more precise statement of the above physical interpretation, in terms of Lagrangian particle trajectories.


\subsection{Equivalence with stochastic Lagrangian approach}
\label{sec:stochastic}

In the forward Lagrangian description \citep{ConstantinIyer08}, a particle position $\widetilde{{\boldsymbol X}}_s^t({\boldsymbol a})$ depends on time $t$ and the particle label ${\boldsymbol a}$, where the label can be defined in terms of the initial position $\widetilde{{\boldsymbol X}}_s^s({\boldsymbol a}) = {\boldsymbol a}$ at the initial time $s$.  
In the backward time approach \citep{Constantin2011}, which is our current interest, one utilizes instead the back-to-label map $\widetilde{{\boldsymbol A}}_T^t({\boldsymbol x}_f)$ which starts from the terminal particle position ${\boldsymbol x}_f$ at the terminal time $T$, and traces back to its original label ${\boldsymbol a}$ at the initial time $s$.  The forward and inverse satisfy $\widetilde{{\boldsymbol X}}_t^T\circ \widetilde{{\boldsymbol A}}_T^t={\rm Id}$, 
where {\rm Id} is the identity map. In presence of viscosity, this back-to-label map can be computed by solving the backward It${\bar{\rm o}}$ SDE for stochastic Lagrangian particle trajectories,
\begin{equation} 
    \hat{d}\widetilde{{\boldsymbol A}}_T^t({\boldsymbol x}_f)={\boldsymbol u}(\widetilde{{\boldsymbol A}}_T^t({\boldsymbol x}_f),t)dt 
    +\sqrt{2\nu}\,\hat{d}\widetilde{{\boldsymbol W}}(t), \quad  s<t<T; \qquad \widetilde{{\boldsymbol A}}_T^T({\boldsymbol x}_f)
    ={\boldsymbol x}_f, 
\label{back_eq} 
\end{equation}
with $dt<0$ and where $\widetilde{\boldsymbol W}$ is Brownian motion.  A few such trajectories are plotted in figure \ref{fig:schematic}. 
Along these trajectories, a material element undergoes tilting and stretching which can be encoded in the deformation matrix, 
$$\widetilde{{\boldsymbol D}}_T^t({\boldsymbol x}_f):=\left. ({\boldsymbol \nabla_a}\widetilde{{\boldsymbol X}}^T_t({\boldsymbol a}))^\top \right|_{{\boldsymbol a}=\widetilde{{\boldsymbol A}}_T^t({\boldsymbol x}_f)}
=({\boldsymbol \nabla_{x_f}}\widetilde{{\boldsymbol A}}_T^t({\boldsymbol x}_f))^{-\top},$$ 
and is computed by solving the ODE, 
\begin{equation} 
    d\widetilde{{\boldsymbol D}}_T^t({\boldsymbol x}_f)=-\widetilde{{\boldsymbol D}}_T^t({\boldsymbol x}_f)    \cdot (\boldsymbol{\nabla u})^\top(\widetilde{{\boldsymbol A}}_T^t({\boldsymbol x}_f),t)dt, \quad s<t<T 
\label{Def_eq} \end{equation} 
backward in time. 
We now have all the ingredients to relate vorticity at the terminal position and time $({\boldsymbol x}_f,T)$ to its back-in-time origin,  
\begin{equation} 
    {\boldsymbol \omega}({\boldsymbol x}_f,T)
        ={\mathbb E}\left[\widetilde{{\boldsymbol D}}_T^s({\boldsymbol x}_f)  {\boldsymbol \omega}(\widetilde{{\boldsymbol A}}_T^s({\boldsymbol x}_f),s)\right]. 
\label{omega_rep2} 
\end{equation} 
The above expression describes the terminal vorticity as the expectation over back-in-time stochastic Lagrangian particles that all emanate from ${\boldsymbol x}_f$ at time $T$.   
It is interesting to note that the above expectation is independent of the initial time $s<T$.
In fact, the random variable in the expectation has been shown to be a backward martingale, or 
statistically conserved quantity backward in time. It has therefore been dubbed the stochastic Cauchy invariant \citep{Eyink2020_theory}, since it generalizes the invariant of \cite{cauchy1815theorie} for Euler solutions to viscous Navier-Stokes solutions.

The stochastic Lagrangian representation using backward evolution in time and the adjoint representation of vorticity can be shown to be exactly equivalent.  We start by rewriting the above expectation as, 
\begin{equation} 
    {\boldsymbol \omega}({\boldsymbol x}_f,T)
        =\int {\mathbb E}\left[\widetilde{{\boldsymbol D}}_T^s({\boldsymbol x}_f) \delta^3\left(\boldsymbol{x}-\widetilde{{\boldsymbol A}}_T^s({\boldsymbol x}_f)\right) {\boldsymbol \omega}(\boldsymbol{x},s)\right] d^3\boldsymbol{x}.
\label{omega_rep2p} 
\end{equation} 
Since the vorticity is no longer a stochastic variable in this representation, we can move it outside the expectation, 
\begin{equation} 
    {\boldsymbol \omega}({\boldsymbol x}_f,T)
        =\int {\mathbb E}\left[\widetilde{{\boldsymbol D}}_T^s({\boldsymbol x}_f) \delta^3\left(\boldsymbol{x}-\widetilde{{\boldsymbol A}}_T^s({\boldsymbol x}_f)\right)\right] {\boldsymbol \omega}(\boldsymbol{x},s) d^3\boldsymbol{x}.  
\label{omega_rep2pp} 
\end{equation} 
Comparing the above form to the duality relation in the adjoint approach, we define the adjoint field using the backward stochastic flow as,
\begin{eqnarray} 
    {\boldsymbol \Upomega}({\boldsymbol x},t) := {\mathbb E}\left[ \widetilde{{\boldsymbol D}}_T^{t}({\boldsymbol x}_f)
    \delta^3({\boldsymbol x}-\widetilde{{\boldsymbol A}}_T^t({\boldsymbol x}_f))\right],
\label{Omega_def} 
\end{eqnarray} 
so that the stochastic representation \eqref{omega_rep2pp} is rewritten as, 
\begin{equation}
    {\boldsymbol \omega}({\boldsymbol x}_f,T)= \int_D {\boldsymbol \Upomega}({\boldsymbol x},s) 
    {\boldsymbol \omega}({\boldsymbol x},s)\, d^3x, \quad s<T.
\label{omega_rep3} \end{equation} 
This expression coincides with the adjoint representation \eqref{eq:operator} derived in the
previous section, in absence of solid boundaries. 

In order to derive the evolution equation for ${\boldsymbol \Upomega}$, we evaluate the time derivative of \eqref{Omega_def}, 
\begin{eqnarray} 
    \partial_t {\boldsymbol \Upomega}({\boldsymbol x},t)dt = {\mathbb E}\left[ \partial_t\widetilde{{\boldsymbol D}}_T^{t}
    \delta^3({\boldsymbol x}-\widetilde{{\boldsymbol A}}_T^t)\right]dt +
    {\mathbb E}\left[ \widetilde{{\boldsymbol D}}_T^{t}
    \hat{d}_t \delta^3({\boldsymbol x}-\widetilde{{\boldsymbol A}}_T^t)\right]. 
\label{eq:dtOmegadt}
\end{eqnarray} 
The first time-derivative $\partial_t\widetilde{{\boldsymbol D}}_T^{t}$ on the right-hand side is given by \eqref{Def_eq}, and the backward It${\bar{\rm o}}$ rule in used for the second term, 
\begin{eqnarray*}
\hat{d}_t \delta^3({\boldsymbol x}-\widetilde{{\boldsymbol A}}_T^t)
= 
-(\hat{d}_t\widetilde{{\boldsymbol A}}_T^t{\boldsymbol \cdot}{\boldsymbol \nabla}_{\boldsymbol{x}}) 
\delta^3({\boldsymbol x}-\widetilde{{\boldsymbol A}}_T^t) 
-\nu\boldsymbol{\nabla}^2_{\boldsymbol{x}} \delta^3({\boldsymbol x}-\widetilde{{\boldsymbol A}}_T^t) dt.
\end{eqnarray*}
Substituting in \eqref{eq:dtOmegadt}, we obtain using 
\eqref{back_eq} and \eqref{Def_eq}  that 
\begin{equation} 
    \partial_t{\boldsymbol \Upomega} 
    = - \boldsymbol{\Upomega}(\boldsymbol{\nabla} \boldsymbol{u})^\top
    -\boldsymbol{u}\cdot\boldsymbol{\nabla}\boldsymbol{\Upomega} - \nu\boldsymbol{\nabla}^2 \boldsymbol{\Upomega}. 
\end{equation}
which coincides with the adjoint vorticity equation \eqref{eq:Omega_eq} after switching to reversed time $\tau=T-t.$

The formula \eqref{Omega_def} is a fundamental result of this section, which provides the direct link between the adjoint and stochastic Lagrangian representations, and gives the precise physical intepretation of the adjoint vorticity field $\boldsymbol{\Upomega}(\boldsymbol{x},t)$, as noted earlier, as the volume density of mean Lagrangian deformation experienced by vorticity from space-time point $(\boldsymbol{x},t)$ to the target point $(\boldsymbol{x}_f,T)$.

The equivalence between the adjoint formulation of section \ref{sec:adjoint} and the stochastic Lagrangian representation applies also in wall-bounded flows.  The Dirichlet and Neuamnn conditions on the adjoint, introduced in the previous section, are equivalent to stochastic counterparts.  In the  Dirichlet case, the stochastic Lagrangian process stops at the wall \citep{Constantin2011, Eyink2020_theory}, while in the Neumann case the stochastic Lagrangian process reflects at the wall \citep{wang_eyink_zaki_2022}.  The details are provided in Appendix \ref{math}.  
This equivalence enhances the importance of both methods. The adjoint vorticity field now 
gains an intuitive Lagrangian interpretation, whereas the stochastic representation gains 
a PDE implementation which is much more computationally efficient than direct Monte Carlo
averaging over Lagrangian trajectories.


\section{Application to turbulent channel flow}
\label{sec:channel}

The utility of the adjoint-vorticity equations \eqref{eq:Omega_eq} is general, as they can be used in conjunction with \eqref{eq:operator} to trace back the origin of vorticity in any viscous, incompressible, free or wall-bounded flow.  
In order to demonstrate this utility, we consider the example of  turbulent channel flow. The flow is periodic in the streamwise ($x$) and spanwise ($z$) directions, and is bounded by no-slip walls at $y\in \{0,2\}$, where lengths are normalized by the channel half height $h^\star$. The velocity scale is the bulk flow speed $U_B^\star$, and the Reynolds number is therefore $Re \equiv U_B^\star h^\star / \nu^\star = 2800$, where $\nu^\star$ is the fluid viscosity.  The corresponding friction Reynolds number is $Re_\tau \equiv u_\tau^\star h^\star / \nu^\star  = 180$, where $u_\tau^\star = \sqrt{\tau^\star_{\textrm{w}}/\rho^\star}$ is the friction velocity and $\tau^\star_{\textrm{w}}$ is the mean wall shear stress and $\rho^\star$ is the density. We emphasize that whenever we use the term ``stress''
in this work we mean the viscous shear stress at the wall and, in particular, its streamwise component associated to drag.  

\begin{figure}
    \centering
    \includegraphics[width=\textwidth]{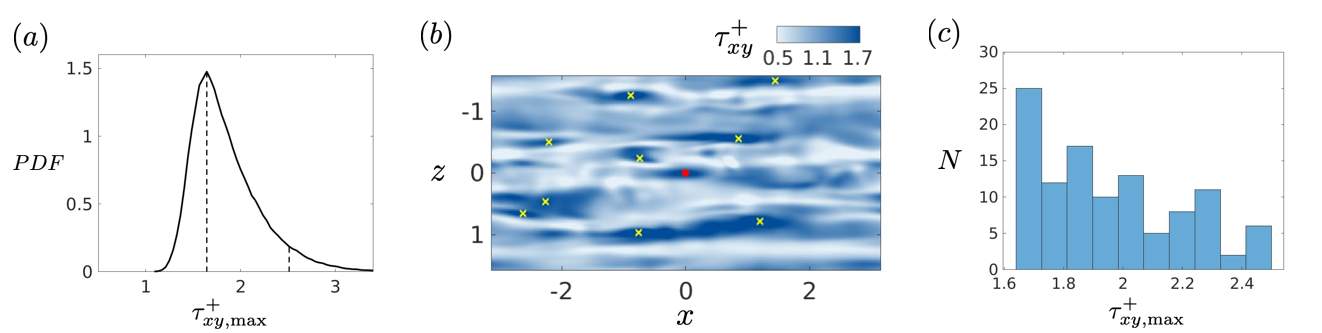}
    \caption{$(a)$ Probability density function (PDF) of all local maxima of the shear stress, during the time horizon $t \in \left[0, 500\right]$. Vertical dashed lines mark the range of events of interest, in the range $\tau^+_{xy,\max} \in \left[1.6, 2.5\right] $.  $(b)$ An instantaneous visualization of the wall shear stress, with uncorrelated local maxima marked by crosses.  $(c)$ Number of uncorrelated events of maximum shear stress on the bottom wall, within the range identified in panel $(b)$.}
    \label{fig:stressPDF}
\end{figure}

\begin{table}
\centering
		\begin{tabular}{c c c c c c c c c c}
			$Re_{\tau}$ & $Re$ & $L_x$& $L_z$& $N_x$ & $N_y$ & $N_z$ & $\Delta x^+$  & $\Delta y^+$ & $\Delta z^+$ \\
			\hline
			180& 2800& $2\pi$& $\pi$& 256& 384& 256& 4.4& 0.36 -- 1.43& 2.2\\
		\end{tabular}
		\caption{Computational domain size and grid parameters.}
		\label{table:grid}
\end{table}

The evolution of the turbulent channel flow is computed using direct numerical simulation. 
The Navier-Stokes equations are advanced using a fractional-step algorithm, where advection terms are treated explicitly using Adams-Bashforth and diffusion terms are treated implicitly using Crank-Nicolson.  The spatial discretization adopts a local volume flux formulation on a staggered grid \citep{Rosenfeld1991}.  The pressure Poisson equation is solved by performing Fourier transforms in the horizontal directions and tridiagonal inversion in the wall-normal direction. 
The algorithm has been extensively tested and applied widely for direct numerical simulations (DNS) of transitional and turbulent flows \citep{zaki2013,Lee_Sung_Zaki_2017}. Validation for channel flow was performed against the DNS data by \citep{Moin_1987}, at the same Reynolds number adopted here, and was presented in \citep{Jelly2014}.
The velocity field is stored throughout the flow evolution, since it is needed for the adjoint equation \eqref{eq:advort_eq}.  The forward vorticity is computed from the stored velocity fields when and where needed.  
Specifically, when decomposing the vorticity at a target space-time point $\left(\boldsymbol{x}_f,T\right)$ into interior and wall contributions from earlier time $t=s$, we require evaluation of the vorticity field $\boldsymbol{\omega}(\boldsymbol{x},s)$ and the wall vorticity or its flux within $t\in[s,T]$.

A discrete adjoint of the Navier-Stokes algorithm is available, which satisfies forward-adjoint duality for the velocity field to machine precision, and which has been adopted for data assimilation in transitional and turbulent flows \citep{wang2019discrete,wang_wang_zaki_2022,wang2021state}. We adopted the same discretization for the numerical solution of the adjoint vorticity equation \eqref{eq:advort_eq}. 
The only new term is the adjoint stretching $\boldsymbol{\nabla} \boldsymbol{u} \cdot \boldsymbol{\Omega}$. We discretized $\boldsymbol{\nabla} \boldsymbol{u}$ using the coordinate-free definition 
$\boldsymbol{\nabla} \boldsymbol{u}=\frac{1}{V_c}\oint_{S_c} d\boldsymbol{S}_c \boldsymbol{u} $ evaluated over the cell volume $V_c$ with bounding areas $\boldsymbol{S}_c$, and we evaluated $\boldsymbol{\nabla} \boldsymbol{u} \cdot \boldsymbol{\Omega}$ at cell centers.  
Our choice for the discretization of the adjoint vorticity does not garner any benefit since it is not the dual of the effective forward vorticity equation.  In this respect, it should be viewed similarly to a continuous adjoint approach.   In order to ensure accuracy, and more specifically that our forward-adjoint duality for the vorticity equation \eqref{eq:dualityk} is satisfied, we have adopted a fine simulation grid (see Table \ref{table:grid}).  In all the results presented herein, the forward-adjoint duality relation \eqref{eq:dualityk} is satisfied to within less than one percent error over the time horizon of interest.

We will focus our analysis on high-stress events on the channel wall, and track their origin in backward time.  We initially identified all the stress maxima during the interval, $t\in\left[0, 500\right]$.  A probability density function (PDF) of these events is shown in figure \ref{fig:stressPDF}$(a)$.  The two marked vertical lines identify the peak of the PDF and an upper bound on the events that we will examine, so we are not including infrequent extreme events in our analysis.  
Within this range, we only considered uncorrelated wall-stress maxima, defined to have a separation in space and time that satisfies $\Delta x\geqslant 1$ or $\Delta z \geqslant 0.15$ or  $\Delta t \geqslant 1$. These criteria were based on the streamwise and spanwise two-point correlations of $\tau_{xy}$ reducing to one half, and the time interval is based on the streamwise separation and the phase speed of the shear stress.
An instance of the wall-stress contours is shown in figure \ref{fig:stressPDF}$(b)$, with the uncorrelated stress events marked by symbols.  The red dot is a particular event that will be analyzed in detail below.  Using these criteria, the total number of uncorrelated stress maxima is 109, and their distribution is reported in figure \ref{fig:stressPDF}$(c)$.

\subsection{Comparison of Dirichlet and Neumann conditions}\label{sec:compare}

For our first demonstration of the back-in-time tracking of vorticity using the adjoint, we will examine a single stress maxima and then proceed to report results from the ensemble of 109 cases.  In order to compare the choice of Dirichlet versus Neumann boundary conditions on the adjoint variable, we must consider a point within the bulk of the fluid since only the Neumann condition is compatible with the initial condition $\boldsymbol{\Omega}^k\left(\boldsymbol{x},T\right) = \boldsymbol{e}_k \delta\left(\boldsymbol{x}-\boldsymbol{x}_{f}\right)$ sampling a point on the wall.  
We therefore initialize the adjoint variable above the location of the stress maximum, at a height $y^+_f=5$.   The particle stress maximum that we consider here is marked by the red circle in figure \ref{fig:stressPDF}($b$).

\begin{figure}
    \centering
    \includegraphics[width=0.9\textwidth]{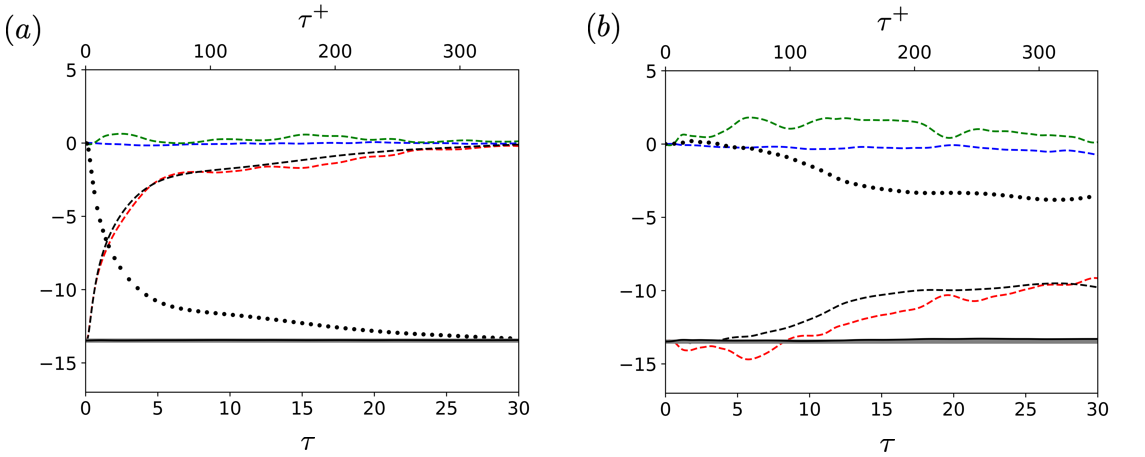}
    \caption{Back-in-time contributions to the target spanwise vorticity $\omega_z^* = \omega_z\left(\boldsymbol{x}_f, T\right)$ for $(a)$ Dirichlet and $(b)$ Neumann boundary condition on the adjoint.   The starting location is at $y_f^+=5$, above the wall-stress maximum shown in figure \ref{fig:stressPDF}$(b)$ marked by a red circle. 
    ({\color{black}$\lline$})
            The total vorticity is evaluated from (\ref{eq:operator}) and compared to ($\graythickline$) the reference value $\omega_z^*$. 
    Internal and boundary contributions:
    {\color{blue}$\dashed$}     $\mathcal{I}_x^z$;
    {\color{green}$\dashed$}    $\mathcal{I}_y^z$; 
    {\color{red}$\dashed$}      $\mathcal{I}_z^z$;
    {\color{black}$\dashed$}    $\mathcal{I}^z$; 
    {\color{black}$\dotted$}    $\mathcal{B}^z$.} 
    \label{fig:duality}
\end{figure}

In figure \ref{fig:duality}, we report the contributions to the dominant spanwise-vorticity at the target point, by evaluating the terms in the integral \eqref{eq:dualityk} with $k=z$, 
\begin{eqnarray}
\begin{aligned}
	\omega_z\left(\boldsymbol{x}_f,T \right)  =& \int_D \boldsymbol{\Omega}^z\left(\boldsymbol{x},s\right) \cdot \boldsymbol{\omega}\left(\boldsymbol{x},s\right) dV ~~~  \\
         & + \int_s^T \oint_{\partial D} 
         \nu  \left(   \left( \boldsymbol{n}\cdot \boldsymbol{\nabla} \boldsymbol{\omega} \right)  \cdot \boldsymbol{\Omega}^z
         	     - \left( \boldsymbol{n}\cdot \boldsymbol{\nabla} \boldsymbol{\Omega}^z\right) \cdot \boldsymbol{\omega} \right) ~dS dt. 
\end{aligned}
\end{eqnarray}
The time horizon is chosen to be $T=30$ (or $T^+=347$), which is equivalent to 1.92 large-eddy turnover times $h/u_{\tau}$. The two panels correspond to the Dirichlet and Neumann boundary conditions.  
The horizontal grey line marks the value of the target spanwise vorticity at $\omega_z(\boldsymbol{x}_f,T) = -13.5$.  The black solid line shows the sum of the right-hand side terms, plotted as a function of the reverse time $\tau=T-t$ preceding the event. The constancy of this value is a consequence of adjoint duality, 
and the observed horizontal line demonstrates the accuracy of our numerical method. 

In addition, however, the reconstruction by the above formula gives precise, detailed information about the origin of the vorticity. In figure \ref{fig:duality} we have divided the target spanwise vorticity into three interior contributions:
\begin{equation}
    \mathcal{I}^z_i(s) = \int_D \Omega_i^z(\boldsymbol{x},s) \, \omega_i(\boldsymbol{x},s)\, dV,
    \qquad i=x,y,z    
\label{eq:intdef} \end{equation}
that result from twisting/tilting of vorticity from the $i=x,y$ directions and stretching of vorticity in the $i=z$ direction, at time $s<T$.  We also report in the figures the boundary contribution for Dirichlet ($D$) or Neumann ($N$) conditions: 
\begin{eqnarray} 
    \mathcal{B}_D^z(s)
    &= -\int_s^T\oint_{\partial D} \nu\frac{\partial\boldsymbol{\Omega}^z}{\partial n}\cdot\boldsymbol{\omega}~dS~dt 
    &=  \int_s^T\oint_{\partial D} B_D^z(s) ~dS~dt  
    \label{eq:BndryD_Int}
    \\
    \mathcal{B}_N^z(s) 
    &=\int_s^T\oint_{\partial D} \nu\frac{\partial\boldsymbol{\omega}}{\partial n} \cdot \boldsymbol{\Omega}^z ~dS~dt,
    &=\int_s^T\oint_{\partial D} B_{N}^z(s)  ~dS~dt
    \label{eq:BndryN_Int}
\label{eq:bdrydef} 
\end{eqnarray} 
arising from either the wall vorticity or its flux, respectively, over the considered time interval $[s,T]$.  
To indicate partial integrals of any quantity $\mathcal{C}$ over variables $A\in\{x,z,t\}$, we will adopt the notation $\overline{\mathcal{C}}^{A}$, etc.

The various contributions for the Dirichlet boundary condition are plotted in figure \ref{fig:duality}$(a)$. 
There is a very substantial increase of the wall contribution over approximately twenty viscous time units.  This back-in-time abrupt return to the wall corresponds in the forward evolution to ``abrupt lifting" \citep{Sheng2009}. However, this process contributes only about 50\% of the target vorticity. The rest arises from the wall only after a much longer interval of several hundred viscous time units, corresponding to a slow diffusion process. The interior contribution at intermediate times arises almost entirely from  $\Omega_z^z,$ indicated by the red dashed line, which corresponds to the spanwise stretching of pre-existing spanwise vorticity.  By contrast, tilting and stretching make negligible contributions. These results are consistent with the high-stress event analyzed by \cite{Eyink2020_channel} numerically using the stochastic Lagrangian approach. However, with the Monte Carlo algorithm of that earlier study, conservation within a few percent was possible for only about 100 viscous time units, even averaging over $N=10^7$ sample paths. In addition, the number of samples required for accurate reconstruction grew exponentially in backward time, so that integrating further was prohibitively expensive.
Here we obtain higher accuracy further back in time for much lower computational cost.  

The various contributions in the case of Neumann boundary conditions are plotted in figure \ref{fig:duality}$(b)$.  The contribution from the wall is now much less significant, and grows very slowly and non-monotonically. Even after $300$ viscous time units, the wall contributes only 26\% of the total.  Most importantly, the dominant contribution arises from interior vorticity. 
Just as for the Dirichlet case, this interior contribution arises very predominantly from  $\Omega_z^z,$ which corresponds to the spanwise stretching of pre-existing spanwise vorticity. A key difference, however, is that spanwise stretching now accounts for the majority of the total vorticity rather than the wall term.  It is perhaps important to underscore that this result is the first of its kind in two ways. Firstly, earlier analysis of the Neumann condition using the stochastic Lagrangian approach has only been attempted in a transitional boundary layer, in order to determine the origin of skin-friction increase at the onset of turbulence spots \cite{wang_eyink_zaki_2022}, and has never been applied in the fully turbulent regime.  Secondly, the importance of the spanwise stretching on internal vorticity is precisely the mechanism suggested by \cite{Lighthill} for concentration and magnification of spanwise vorticity in near-wall turbulence. Whether this behaviour is statistically typical, beyond this single analyzed event, is a question that we will address later in this section.

\begin{figure}
    \centering
    \includegraphics[width=\textwidth]{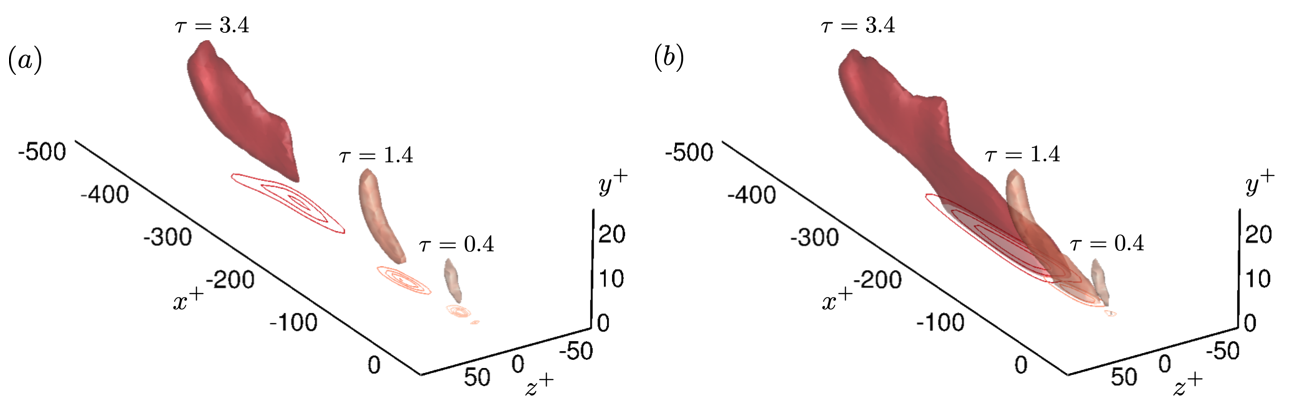}
    \caption{Iso-surfaces of $\Omega_{z}^z/\max |\Omega_{z}^{z}|$ as a function of backward time at $\tau=\{0.4, 1.4, 3.4\}$ for $(a)$ Dirichelet and $(b)$ Neumann boundary conditions. The line contours are the wall values for $(a)$ $\nu \left(\partial \Omega_{z}^z / \partial y\right)$ and $(b)$ $\Omega_{z}^z$. The vertical axes are stretched for clarity of the visualization.}
    \label{fig:adjoint_iso}
\end{figure}        

For a detailed view of the space-time origin of vorticity, we plot 
in figure \ref{fig:adjoint_iso} iso-surfaces of the adjoint field $\Omega_z^z$ and iso-contours of the wall terms ($\nu\partial \Omega_z^z/\partial y$ for Dirichlet, $\Omega_z^z$ for Neumann). 
The figure shows three time instances together, $\tau=\{0.4, 1.4, 3.4\}$,
which are selected to be within the interval of fast changes in the contributions in the Dirichlet case (see figure \ref{fig:duality}). 
The expanding iso-surfaces in backward time represent the spreading of the adjoint field, which accounts for the accumulated stretching rate of earlier vorticity as it is transported by advection and viscous diffusion.  
The level sets of the wall terms clearly lag behind those of the interior. This observation is easily understood by the faster streamwise fluid velocities at greater distances from the wall. 
The most prominent difference between the two figures is related to the interior contribution.  While the iso-surfaces appear similar away from the wall, they are qualitatively different in the near-wall region.  Specifically, the iso-surfaces of the Neumann adjoint have a large lobe near the wall, which is entirely missing for the Dirichlet case. 
This difference is intuitive since the Neumann condition is akin to an adiabatic condition while the Dirichlet counterpart annihilates the adjoint at the wall.  In the stochastic formulation, the former condition reflects the particles when they hit the wall, whereas the latter absorbs the particles and hence depletes the adjoint field.

\begin{figure}
            \centering
            \includegraphics[width=0.9\textwidth]{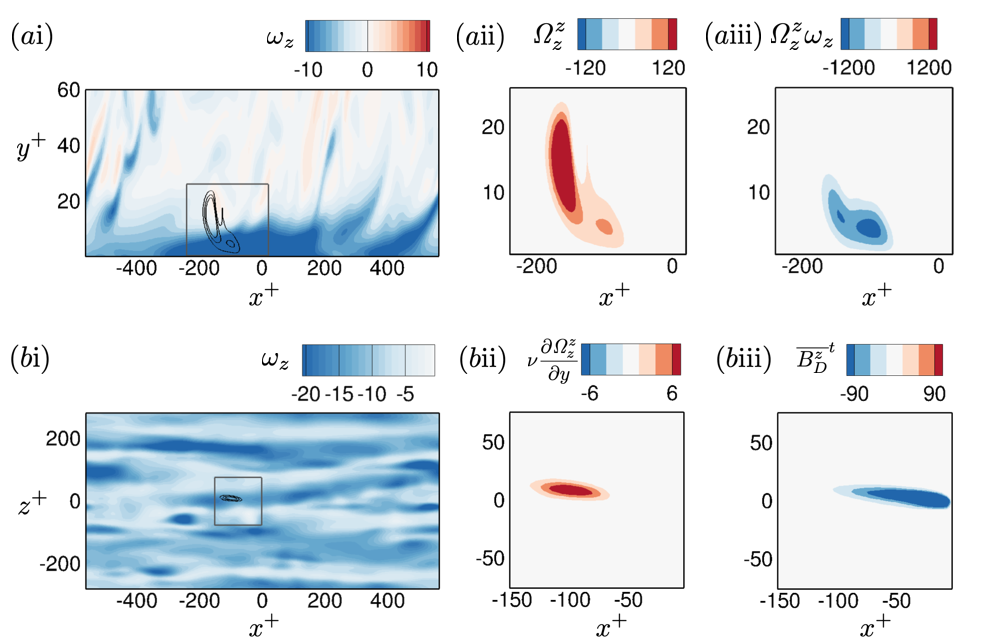}
            \caption{Contributions to the terminal vorticity at $\tau=1.4$, for Dirichlet boundary conditions. 
            $(a\mathrm{i})$ Side view showing color contours of the spanwise vorticity and lines contours of the adjoint $\Omega_{z}^{z}$.
            $(a\mathrm{ii})$ Zoomed-in view with colors showing $\Omega_{z}^{z}$.  
            $(a\mathrm{iii})$ The contribution to the terminal vorticity by spanwise stretching $\Omega_{z}^{z}\omega_z$. These side-view plots are all vertically stretched.
            $(b\mathrm{i})$ Top view of the wall, showing color contours of the spanwise vorticity and lines contours of $\nu \partial \Omega_{z}^{z} / \partial y$.
            $(b\mathrm{ii})$ Zoomed-in view with colors showing $\nu \partial \Omega_{z}^{z} / \partial y$.              $(b\mathrm{iii})$ The boundary contribution $\overline{B_D^z}^t$.}
             \label{fig:dirichlet_tau1}

            \vspace*{12pt}
            \centering
            \includegraphics[width=0.9\textwidth]{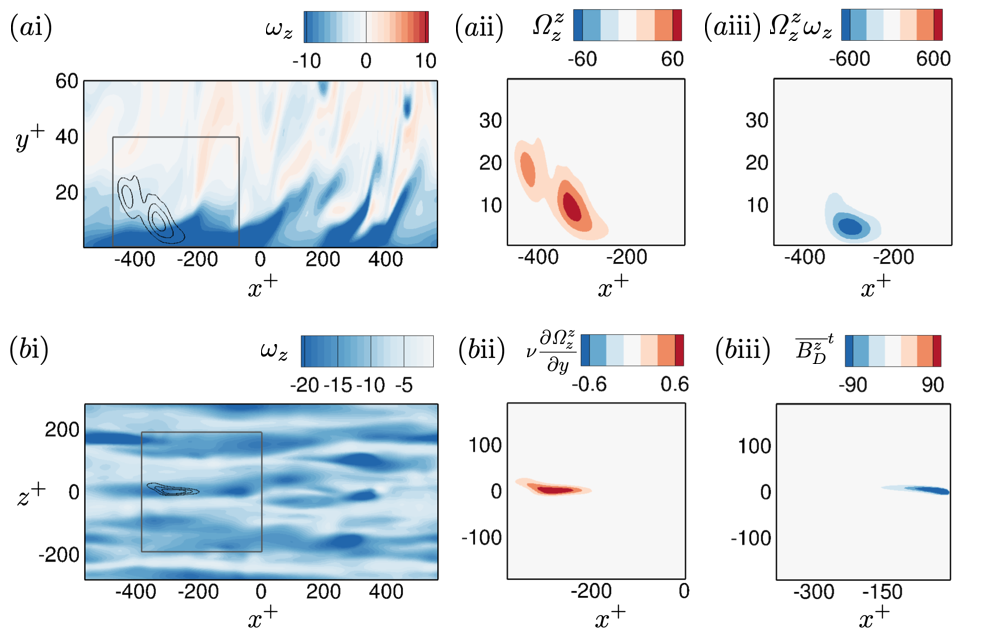}
            \caption{
            Same as figure \ref{fig:dirichlet_tau1}, except showing the contributions to terminal vorticity at $\tau=3.4$, and the contours levels are adjusted as marked.}
            \label{fig:dirichlet_tau2}
        \end{figure}

\begin{figure}
            \centering
            \includegraphics[width=0.9\textwidth]{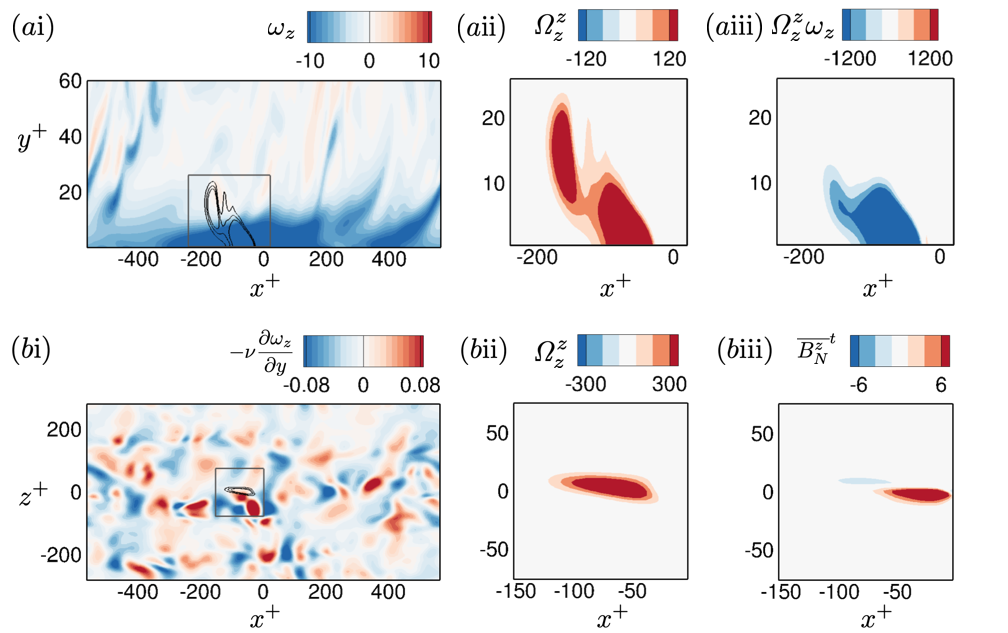}
            \caption{Contributions to the terminal vorticity at $\tau=1.4$, for Neumann boundary conditions. 
            $(a\mathrm{i})$ Side view showing color contours of the spanwise vorticity and lines contours of the adjoint $\Omega_{z}^{z}$. 
            $(a\mathrm{ii})$ Zoomed-in view with colors showing $\Omega_{z}^{z}$.  
            $(a\mathrm{iii})$ The contribution to the terminal vorticity by spanwise stretching $\Omega_{z}^{z}\omega_z$. These side-view plots are all vertically stretched.
            $(b\mathrm{i})$ Top view of the wall, showing color contours of $-\nu \partial \omega_{z} / \partial y$ and lines contours of $\Omega_{z}^{z}$.  
            $(b\mathrm{ii})$ Zoomed-in view with colors showing $\Omega_{z}^{z}$.  
            $(b\mathrm{iii})$ The boundary contribution $\overline{B_N^z}^t$.}
            \label{fig:neumann_tau1}

            \vspace*{12pt}
            \centering
            \includegraphics[width=0.9\textwidth]{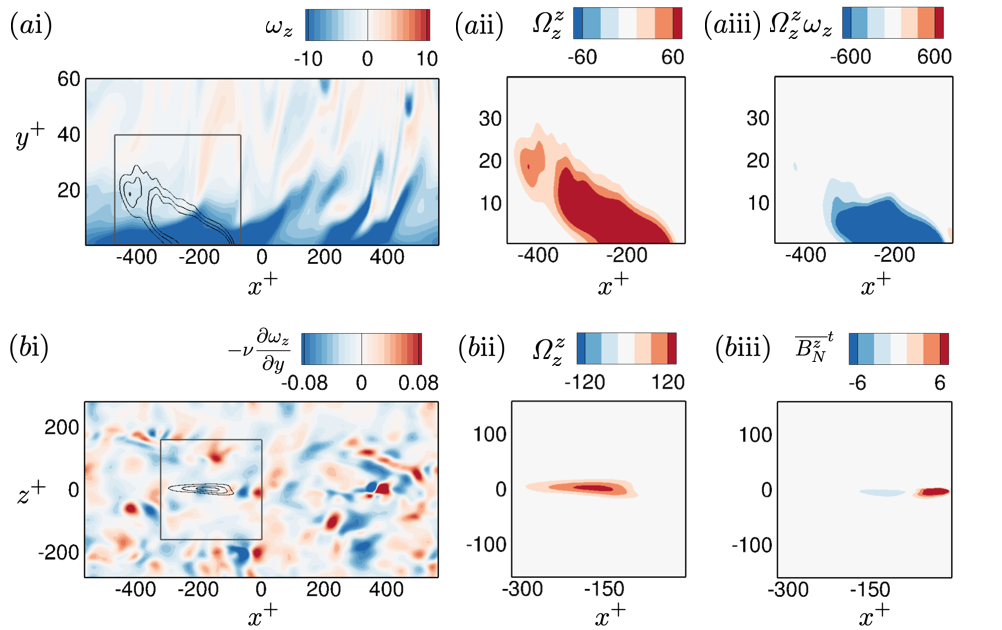}
            \caption{
           Same as Figure \ref{fig:neumann_tau1}, except showing the contributions to the terminal vorticity at $\tau=3.4$, and the contours levels are adjusted as marked.}
            \label{fig:neumann_tau2}
        \end{figure}

\begin{figure}
            \centering
            \includegraphics[width=0.9\textwidth]{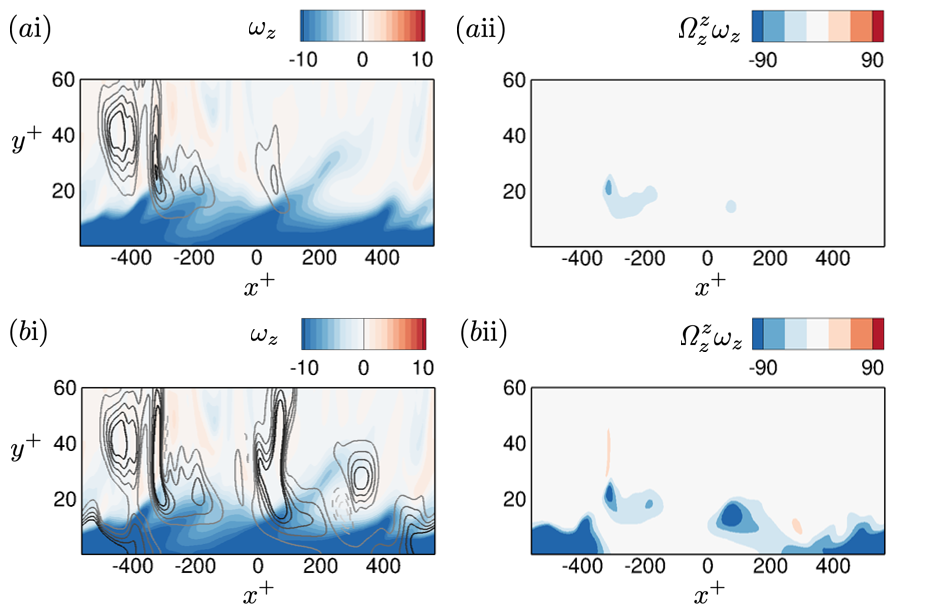}
            \caption{Comparison of the internal contributions to the terminal vorticity at the large backward time $\tau = 10$, using $(a)$ Dirichlet and $(b)$ Neumann boundary conditions. 
            (i) Side view showing contours of spanwise vorticity, overlaid by lines of the adjoint $\Omega_{z}^{z}$. Solid and dashed line contours represent positive and negative values of $\Omega_z^z$, respectively.  (ii) Contours of the contribution of vorticity stretching $\Omega_{z}^{z} \omega_{z}$.  The plane is located at the spanwise location where the target vorticity is sampled. These side-view plots are all vertically stretched.}
            \label{fig:compare_tau3}
        \end{figure}

The total contribution to the target vorticity arises, however, not only from the adjoint field but also from its 
products with the interior vorticity and with the wall vorticity or its flux. These initial vorticities and 
wall conditions are advected and diffused, stretched and rotated, to the target point, where 
they are fused by viscosity to yield the resultant vorticity. These fluid dynamical processes are all precisely represented by the adjoint vorticity, or density of the deformation field. The next several figures \ref{fig:dirichlet_tau1}-\ref{fig:compare_tau3} illustrate how these processes operate in the specific high wall-stress event discussed thus far.


The Dirichlet case is shown first in figures \ref{fig:dirichlet_tau1}-\ref{fig:dirichlet_tau2}, for the two backward times $\tau=1.4$ and $\tau=3.4,$ respectively.  
The target point is at $y^+=5$, and its horizontal coordinates are taken as the origin in the the wall $xz$-plane.  
The top rows $(a)$ of both figures focus on the interior contribution to the target vorticity.  
The three panels show a side view at the target location: $(a\mathrm{i})$ color contours of the spanwise vorticity $\omega_z$ and lines for $\Omega_z^z$; $(a\mathrm{ii})$ a zoomed-in view of $\Omega_z^z$; and $(a\mathrm{iii})$ their instantaneous product $\Omega_z^z\omega_z$.  The last term, once integrated over volume, is the largest interior contribution to the target vorticity even if it is decaying while the wall value is increasing. 
To interpret these results, it is helpful to recall that $\Omega_z^z$ is the volume density of mean spanwise stretching.  The color contours in $(\mathrm{ii})$ for the 
two times show clearly that the adjoint backward in time samples regions further upstream and vertically higher. The vorticity $\omega_z,$ however, is strongly concentrated near the wall, with narrow tongues ejected into the interior. Thus, the product $\Omega_z^z\omega_z$ arises also mainly from the near-wall region, although the contours are attenuated because of the Dirichlet condition on $\Omega_z^z$.  

The bottom rows $(b)$ of figures \ref{fig:dirichlet_tau1}-\ref{fig:dirichlet_tau2} relate to the contribution from the wall vorticity.  
The three panels show a top view in the wall plane: $(b\mathrm{i})$ color contours of the spanwise vorticity $\omega_z$ and lines for $\nu\partial \Omega_z^z/\partial y$; $(b\mathrm{ii})$ a zoomed-in view of $\nu\partial \Omega_z^z/\partial y$; and $(b\mathrm{iii})$ the time integral of their product over the interval $[T-\tau,T]$ which represents the cumulative contribution  to the target vorticity per area.  
Recall that the quantity in panel $(b\mathrm{ii})$ represents the density per area and time of mean deformation of wall vorticity, which ultimately reaches the target location.  
The figure clearly illustrates that the adjoint samples the wall farther upstream backward in time. However, the product field which is the boundary contribution shown in $(b\mathrm{iii})$ is retarded relative to the adjoint field and to the interior contribution in $(a\mathrm{iii})$, because it is cumulative and is dominated by early times when the adjoint field first reaches the wall.

The analogous plots in figures \ref{fig:neumann_tau1}-\ref{fig:neumann_tau2} for the Neumann case show the same event and the same times, for direct comparison. The top rows $(a)$ in the two new figures show the analogous interior quantities.  
The interior contributions may appear similar to those for Dirichlet boundary conditions except, as already noted, there is a much larger near-wall region due to the adiabatic condition.  As a result, the interior contribution for the Neumman case is much larger, and is in fact the overall dominant term as reported in the overall duality balance (see figure \ref{fig:duality}).
The boundary contribution in the Neumann case is particularly interesting since it is related to the vorticity flux at the wall.  The relevant quantities are reported in  the bottom rows of figures \ref{fig:neumann_tau1}-\ref{fig:neumann_tau2}:  $(b\mathrm{i})$ color contours of the Lighthill vorticity source $(-\nu\partial \omega_z/\partial y)$ and lines for the adjoint $\Omega_z^z$; $(b\mathrm{ii})$ a zoomed-in view of $\Omega_z^z$; and $(b\mathrm{iii})$ the time integral of their product over the interval $[T-\tau,T]$, which again represents the cumulative contribution to the target vorticity per area.  
The Lighthill source in panels $(b\mathrm{i})$ is notable in having values of both positive  and negative sign, distinctly different from the spanwise vorticity itself which is almost always the same sign as the mean vorticity. 
Since at the wall $(-\nu\partial \omega_z/\partial y)=\partial p/\partial x$, this quantity has a negative mean value in the pressure-driven channel flow.  However, this mean is more than an order of magnitude smaller than the fluctuations, and its relative magnitude decreases $\propto 1/Re_\tau$.  This very small mean pressure gradient was already argued by \cite{Lighthill} to be unable to account for the strong near-wall concentration of spanwise vorticity, writing that ``even in an accelerating flow, the rate of production ... is too small'' (p.98). The argument by \cite{Lighthill} is not entirely compelling, however, because it does not take account of the very large fluctuating values of the pressure gradient. 
Nevertheless, panels ($b\mathrm{i}$-$\mathrm{ii}$) show that the adjoint samples vorticity sources of both signs and the cumulative contribution may suffer extensive cancellation, and hence the small total boundary contribution in figure \ref{fig:duality} at early times.   
Notice, in fact, that for this particular example the boundary contributions at both times, $\tau=\{1.4, 3.4\}$, have net positive sign which is opposite to the target vorticity.

        \begin{figure}
            \centering
            \includegraphics[width=0.9\textwidth]{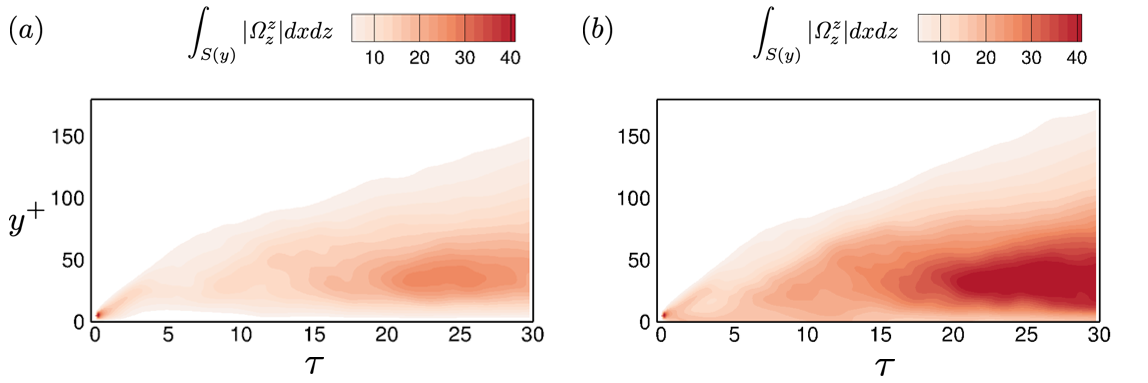}
            \caption{Comparison of horizontally integrated $|\Omega_{z}^{z}|$ as a function of backward time, for $(a)$ Dirichlet and $(b)$ Neumann boundary conditions.}
            \label{fig:int_Omega}
        \end{figure}       

In figure \ref{fig:compare_tau3}, we directly compare the interior contributions using Dirichlet and Neumann boundary conditions for the same event, at an earlier time $\tau=10$.   At this time, the target vorticity in the Dirichlet case is almost entirely from the wall, while for the Neumann case the majority contribution remains due to the interior. 
In subpanels (i) of these figures, the spanwise vorticities $\omega_z$ are the same, but the level sets of $\Omega_z^z$, while very similar far from the wall, show that the Neumann adjoint field samples the near-wall region much more densely.  Because this region is also the site of the strongest magnitudes of $\omega_z$, the volumetric contributions $\Omega_z^z\omega_z$ plotted in the panels (ii) shows much larger values for the Neumann case and is heavily localized near the wall. 

Returning to figures \ref{fig:compare_tau3}(i), we may note that, at these early times, some of the contours of $\Omega_z^z$ are negative, which can be explained by adjoint chaos in the buffer layer \citep{wang_wang_zaki_2022}: 
The adjoint vorticity is advected and tilted and stretched by the turbulent velocity field. 
Furthermore, to make our observations regarding the adjoint fields near the wall more quantitative, we plot in figure \ref{fig:int_Omega} the $L^1$-norms $\int_{S(y)}|\Omega_z^z|~dx~dz$ as functions of $y^+$ and backward time $\tau=T-t$. These show a vertical spreading in $y$, for both the Dirichlet and Neumann boundary conditions, as back-in-time advection extends the adjoint field upward.  There is also a strengthening of magnitudes with $\tau$, associated with the accumulated stretching.  Most importantly, while the Dirichlet condition depletes the near-wall adjoint field, the Neumann condition preserves the near-wall magnitudes which corresponds to a high density of stretching in this region.  For this reason, the near-wall stretching of internal vorticity remains the dominant term in the overall makeup of the terminal vorticity (see figure \ref{fig:duality}). 

Thus far, we have considered a particular high-stress event, and examined it in detail using the adjoint fields with Dirichlet and Neumann boundary conditions. Whether the reported trends are statistically robust is examined in the next section.

\subsection{Ensemble analysis of spanwise vorticity above stress maxima} \label{sec:statistical}

In the remainder of this section we analyze the entire ensemble of 109 local stress-maxima that were identified.  In order to continue the comparison of the Dirichlet and Neumann adjoint boundary conditions, we must examine a point in the interior of the fluid domain.  We therefore continue the analysis of the spanwise vorticities at $y^+=5$, above these local wall maxima.  Our aim here is to identify generic features in the development of such strong near-wall vorticities. 
We will then exploit the Neumann condition, which can be applied for the analysis of vorticity points that are directly on the wall, in order to study the origin of  the extrema in the spanwise wall shear stress.  

        \begin{figure}
            \centering
            \includegraphics[width=1\textwidth]{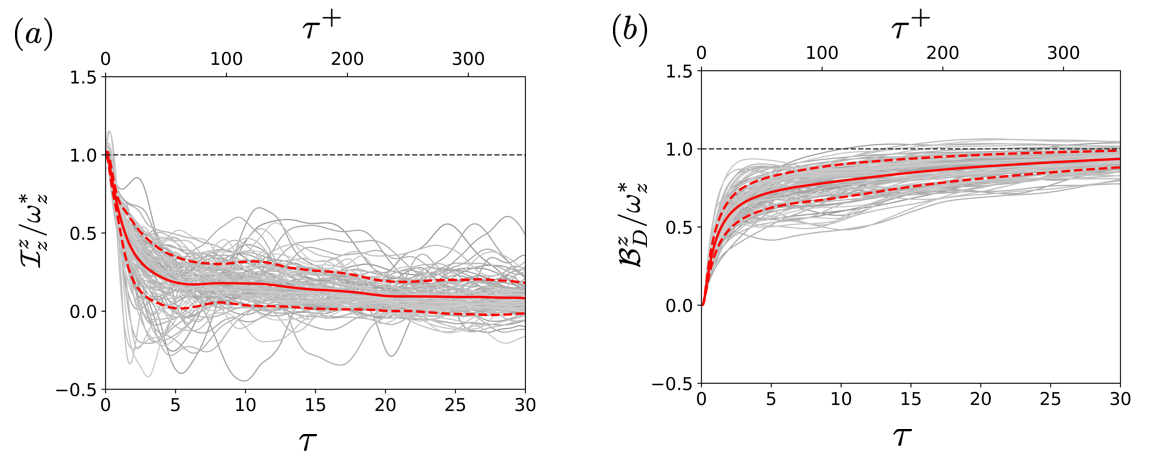}
            \caption{Ensemble of 109 target vorticity events, at $y^+=5$ above a wall-stress maximum, tracked back in time using the Dirichlet condition.  
            Gray lines are the $(a)$ interior vorticity stretching ${\mathcal I}^z_z=\Omega_{z}^{z} \omega_{z}$  and $(b)$ boundary contributions to the target vorticity, normalized by the target values $\omega_z^* = \omega_z\left(\boldsymbol{x}_f, T\right)$.             ({\color{red}$\lline$}) Ensemble-averaged value;             ({\color{red}$\dashed$}) $\pm$ the standard deviation.              }
            \label{fig:EnsembleDirichlet}
        \end{figure}  

        \begin{figure}
            \centering
            \includegraphics[width=1\textwidth]{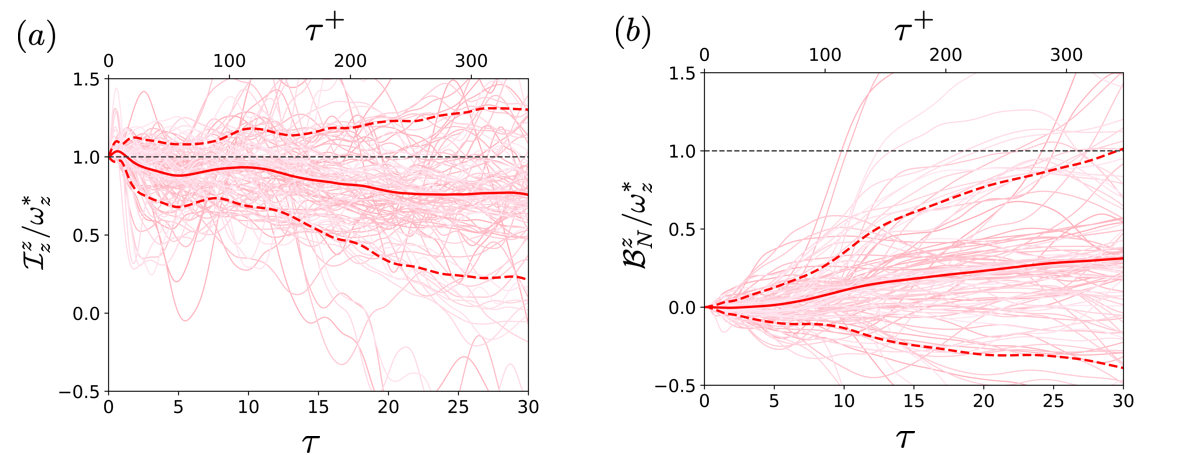}
            \caption{Same as figure \ref{fig:EnsembleDirichlet}, except that the target vorticity is tracked back in time using the Neumann condition. 
            }
            \label{fig:EnsembleNeumann}
        \end{figure}

We begin by plotting in figure~\ref{fig:EnsembleDirichlet}, as a function of reverse time $\tau$, the two largest fractional contributions to the target vorticity when the Dirichlet boundary condition is adopted. 
In figure $(a)$, we report the spanwise stretching of interior spanwise vorticity, $\mathcal{I}_z^z$ as defined in \eqref{eq:intdef} for $i=z$, and in $(b)$ we plot the wall-vorticity contribution, $\mathcal{B}_{D}^z$ as defined in \eqref{eq:BndryD_Int}.  
The light curves are the 109 ensemble members, and the red lines mark the mean plus-and-minus a standard deviation.   
We see that many of the features observed for the results of the single realization plotted in figure~\ref{fig:duality}(a) hold quite generally. The entire contribution arises at very small $\tau$ (immediately before the target time) from $\mathcal{I}_z^z$ and the decline in this contribution in backward  time is compensated by a growth in the contribution from $\mathcal{B}_{D}^z.$ 
There is remarkable similarity in the time histories for all realizations, although with clear fluctuations and occasional large excursions.  In all cases, the contribution from spanwise stretching of interior vorticity declines to near 8\% within a few hundred viscous times and the contribution from wall vorticity rises to near 93\% in the same time. The development of large $\omega_z^*$ at $y^+=5$ appears to occur in a very regular fashion when traced back to wall vorticity, which is demonstrated by the small standard deviation for vorticity stretching and boundary contributions in figure \ref{fig:EnsembleDirichlet}.

The analogous data for the adjoint with Neumann boundary conditions are shown in figure~\ref{fig:EnsembleNeumann}.  We again plot the fractional contributions to $\omega_z^*$ from $(a)$ the interior $\mathcal{I}_z^z$, and from $(b)$ the boundary term  $\mathcal{B}_{N}^z$.  Note that the latter samples the wall vorticity flux, as defined in \eqref{eq:BndryN_Int}.  We see again that many of the features observed for the results of the single realization plotted in figure~\ref{fig:duality}$(b)$ hold quite generally.   The stretching of interior spanwise vorticity, $\mathcal{I}_z^z$, accounts for an appreciable fraction of the total, whether for the individual cases or for the average.  
Different from the specific realization plotted in figure~\ref{fig:duality}, some of the curves of $\mathcal{I}_z^z$ persist in exceeding the target value.  In addition, while the boundary contribution $\mathcal{B}_{N}^z$ generally grows backward in time, some of the events have the opposite sign to the target vorticity.  In other words, the flux of vorticity from the wall reduces the observed value and must be compensated by other terms. 
Of course, this is quite understandable given the space plots of the vorticity source $\sigma_z$ over the wall in figures~\ref{fig:neumann_tau1}-\ref{fig:neumann_tau2}$(b\mathrm{i})$, which reveal that the instantaneous flux takes on large values of both signs.  
We know that the boundary flux contribution must grow to 100\% extremely far back in time since all the vorticity originates ultimately at the walls. However, even at the final time $\tau_2=30$ in figure~\ref{fig:EnsembleNeumann} the mean contribution over the ensemble is only about 31\%.  
We see much larger fluctuations over the ensemble for Neuman boundary conditions than we did in the Dirichlet case, which means that the origin of high $\omega_z$ events at $y^+=5$ appears much more idiosyncratic in terms of wall vorticity flux than it does in terms of wall vorticity. 
This state of affairs is reflected 
in the appreciable increase of the standard deviation of the interior and wall contributions in figure \ref{fig:EnsembleNeumann}. 
At early time $\tau \approx 1$, the figure shows sharp distributions with $\mathcal{I}_z^z/\omega_z^*\simeq 1$ and $\mathcal{B}_{N}^z/\omega_z^*\simeq 0$, whereas at 
$\tau=30$ both distributions have become quite broad, with $\mathcal{I}_z^z$ contributing 75\% of $\omega_z^*$ on average and $\mathcal{B}_{N}^z$ contributing 31\% on average, but with standard deviations of order 54\% for $\mathcal{I}_z^z$ and 70\% for $\mathcal{B}_{N}^z.$  Additional statistical characterization of the results in figures \ref{fig:EnsembleDirichlet} and \ref{fig:EnsembleNeumann} are presented in Appendix \ref{statcontr}.

        \begin{figure}
            \centering
            \includegraphics[width=1\textwidth]{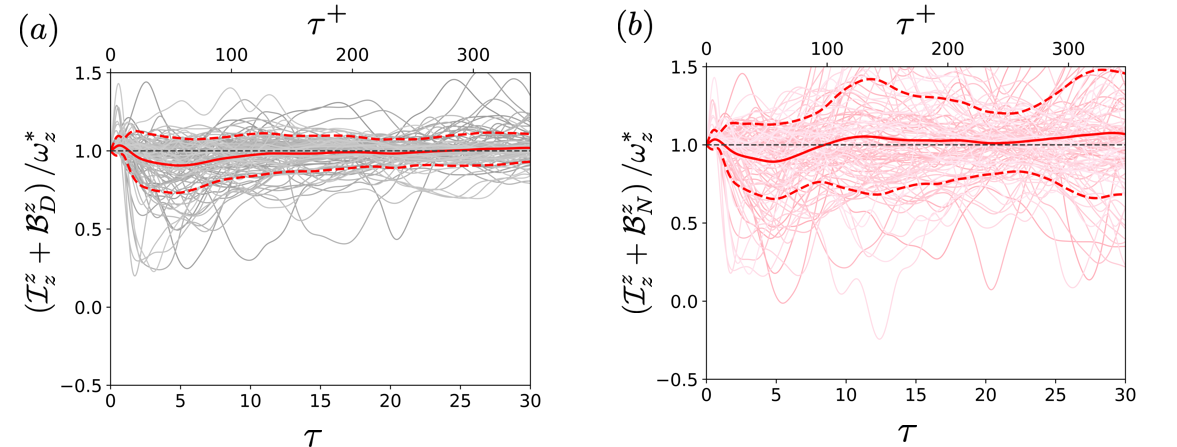}
            \caption{Ensemble of 109 target vorticity events, at $y^+=5$ above a wall-stress maximum, tracked back in time showing the sum $\left(\mathcal{I}_z^z+\mathcal{B}^z_{D/N}\right)/\omega_z^*$ of interior stretching and boundary contributions for $(a)$ Dirichlet $(D)$ and $(b)$ Neumann $(N)$ boundary conditions. 
            }
            \label{fig:EnsembleSum}
        \end{figure}

In figure \ref{fig:EnsembleSum}, we report the sum of the two dominant terms, namely the stretching of interior spanwise vorticity $\mathcal{I}_z^z$ and the boundary contribution $\mathcal{B}^z_{D/N}$.  Figure \ref{fig:EnsembleSum}$(a)$ is for the Dirichlet case, and $(b)$ is the Neumann counterpart.  
We see that the sum of these two terms is close to $\omega_z^*$ for all reverse times, so that we can infer that the other interior terms, $\mathcal{I}_x^z$ and $\mathcal{I}_y^z$, are relatively small, certainly on average.  In addition, for the Neumann case, the variance in the sum is reduced relative to the individual terms (compare figures \ref{fig:EnsembleSum}$b$ and \ref{fig:EnsembleNeumann}).  

In summary, the key contribution to the vorticity at $y^+=5$, above a high wall-stress event is due to the combined effects of stretching on interior vorticity and the boundary.  When the adjoint satisfies the homegeneous Dirichlet condition, it samples the wall vorticity, which progressively becomes the more important of the two contributions in this case ($50\%$ by $\tau^+\approx 20$ and exceeds $70\%$ by $\tau^+ \gtrsim 50$).  However, this choice does not account for the origin of that wall vorticity itself.  In the Neumann case, the dominant contribution within the entire interval of reverse time that we considered ($\tau^+ \le 347$) was primarily due to the spanwise stretching of interior vorticity.  The Neumann condition also has the advantage of being applicable directly at the wall, starting from the high-stress events themselves\textemdash the focus of the subsequent analyses.

\subsection{Ensemble analysis of wall-stress maxima} 
\label{sec:wallstress}

\begin{figure}
            \centering
            \includegraphics[width=0.9\textwidth]{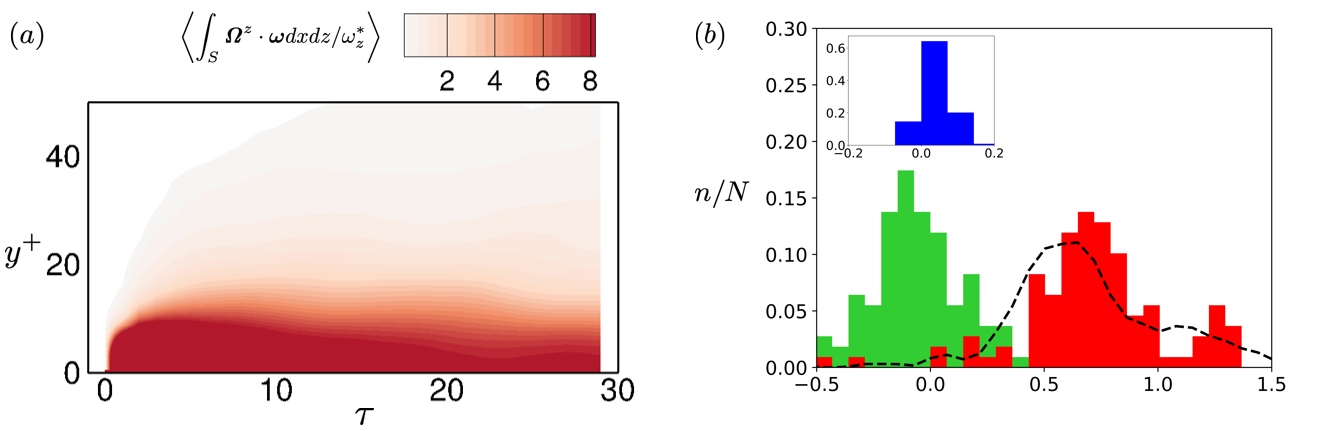}

            \caption{
            $(a)$ Horizontally integrated interior contribution to the wall vorticity at stress maxima, as a function of backward time. The integral is normalized by the target vorticity, and ensemble averaged. 
            $(b)$ Distribution of interior contributions at $\tau=30$, all normalized by the target value $\omega_z^*$.  ($\dashed$) The total interior contribution (dashed line), and its (red) spanwise  $\mathcal{I}_{z}^z$, (green) wall-normal $\mathcal{I}_{y}^z$, and (blue) streamwise 
            $\mathcal{I}_{x}^z$ components.}
            \label{fig:InteriorWall}
         \end{figure}

Starting from each of the 109 high-stress events on the wall, we have tracked back their origin using the adjoint approach with Neuman boundary conditions.  The general trends are similar to the results from $y^+=5$, so we will here report new quantities that were not previously discussed, in order to expand the scope of the analysis.   
In figure \ref{fig:InteriorWall}($a$), we report the horizontally integrated interior contribution, normalized by the target value and ensemble averaged over the 109 events.  The contours are plotted as a function of $y^+$ and reverse time.  The results show that the interior term is concentrated in the near-wall region, $y^+ \lesssim 15$.  This trend is the combined effect of the adjoint vorticity being large near the wall, as we have seen in figure \ref{fig:int_Omega} for example, and the vorticity being largest at the wall.   In figure \ref{fig:InteriorWall}$(b)$, we report the distribution of the interior terms $\mathcal{I}_{(x,y,z)}^z$ and their sum, at $\tau=30$.   The streamwise and wall-normal contributions $\mathcal{I}_{(x,y)}^z$ are, as anticipated, distributed near zero.  The spanwise stretching term is by far the largest, accounting for approximately $75\%$ of the target value.  In addition, this term accounts for the majority of the total interior contribution.

\begin{figure}
            \centering
            \includegraphics[width=0.7\textwidth]{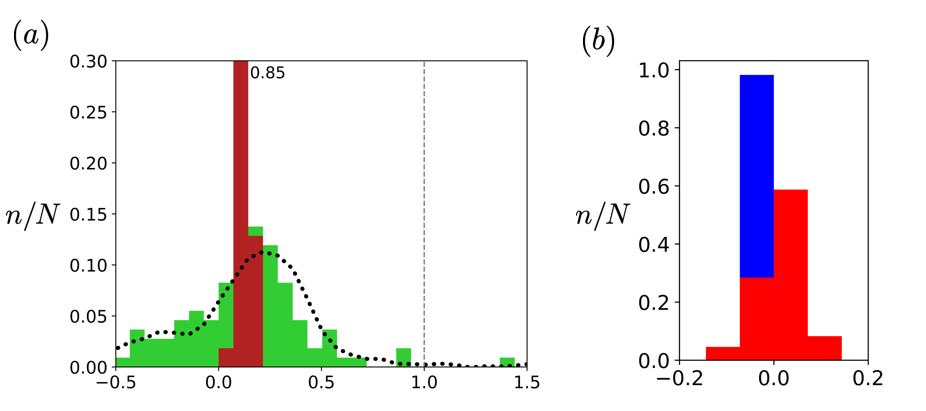}
            \caption{
            Distribution of boundary contributions to the wall vorticity at the stress maxima, at $\tau=30$,
            all normalized by the target value $\omega_z^*$.   
            $(a)$  Total 
            boundary contribution $\mathcal{B}_{N}^z$ (dotted line); contribution 
            $\mathcal{B}_{N\langle z \rangle}^z$ due to the mean streamwise pressure gradient (dark red); 
            and contribution $\mathcal{B}_{Ny}^z$ due to tilting of wall-normal vorticity injected at the boundary (green).  
            $(b)$ Contributions from $\mathcal{B}_{Nz^\prime}^z$ due to fluctuating 
            streamwise pressure-gradient (red) and from $\mathcal{B}_{Nx}^z$ due to twisting of streamwise 
            vorticity produced at the wall (blue).}
            \label{fig:BoundaryWall}
        \end{figure}

The boundary term is examined in figure \ref{fig:BoundaryWall}.  The distribution is of the fractional boundary contributions from the ensemble of events, at $\tau=30$ or $\tau^+=347$.   Interestingly, this distribution has a small positive mean of about $0.25$.   
To test if the small mean value we observe arises entirely from the mean streamwise pressure gradient, we consider the contributions from separate components of the vorticity source,
\begin{equation} 
    \mathcal{B}_{Ni}^z(s) =\int_s^T\oint_{\partial D} \Omega^z_i\sigma_i~dS~dt 
    \qquad    i=x,y,z 
\label{eq:bdrydefnm} 
\end{equation} 
where $\sigma_i=-\nu \partial \omega_i/\partial y$, and furthermore decompose the spanwise term into two parts, 
\begin{equation} 
    \mathcal{B}_{Nz}^z = \mathcal{B}_{N\langle z \rangle}^z + \mathcal{B}_{Nz^\prime}^z, 
\end{equation} 
where $\mathcal{B}_{N\langle z \rangle}^z$ arises from the mean source $\overline{\sigma}_z=\partial \overline{p}/\partial x$ and $\mathcal{B}_{Nz^\prime}^z$ is from the fluctuations $\sigma_z'=\partial p'/\partial x$.  The contribution from the mean pressure gradient does have a small positive net value of about $0.12$, about half of the distribution mean but with a very small variance (which arises solely from the variance of $\Omega_z^z$ averaged over target points).  As such, this term does not account for the observed variance of $\mathcal{B}_{N}^z$ and only 
half of its mean. We have found instead that $\mathcal{B}_{Ny}^z$,
due to the tilting of wall-normal vorticity injected at the wall, accounts for the distribution of the boundary term, as shown by the results plotted in figure~\ref{fig:BoundaryWall}($a$).

\begin{figure}
            \centering
            \includegraphics[width=0.9\textwidth]{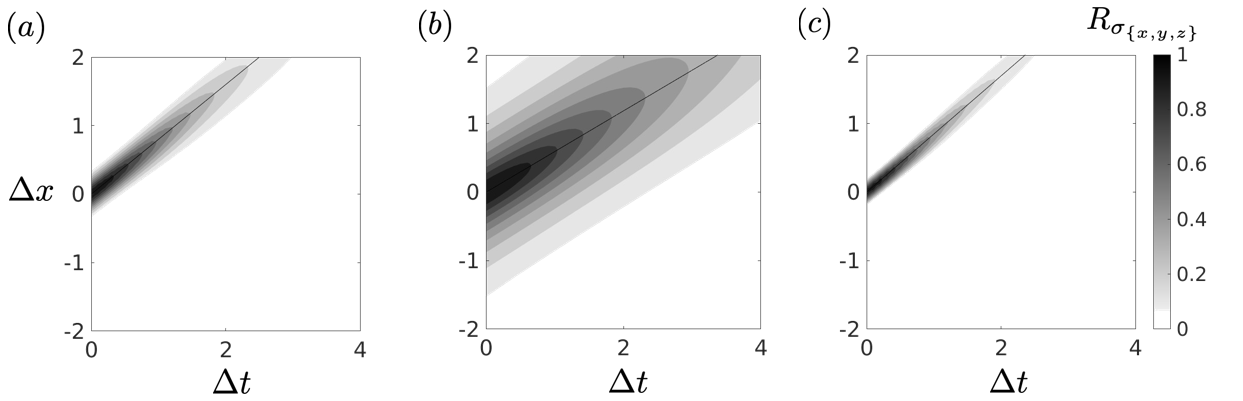}
            \caption{Two-point space-time correlation of the wall-normal flux of $(a)$ streamwise, $(b)$ wall-normal, and $(c)$ spanwise vorticity. The three marked lines correspond to speeds $c=\{0.8, 0.59, 0.84\}$, or equivalently $c/u_\tau = \{12.5, 9.2, 13.1\}$.}
            \label{fig:c_flux}
        \end{figure}

\begin{figure}
            \centering
            \includegraphics[width=0.9\textwidth]{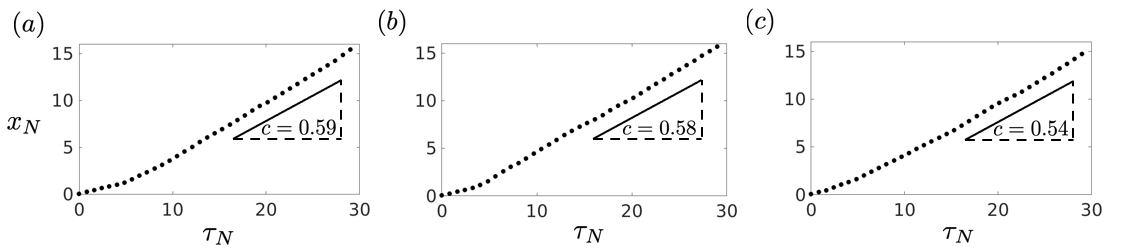}
            \caption{Phase speed for wall adjoint vorticity in its $(a)$ streamwise, $(b)$ wall-normal, and $(c)$ spanwise components. Here $x_{N}=\sum_{n=1}^{N-1}\Delta x_{n}$, where $\Delta x_{N}, \Delta z_{N}=\mathrm{argmax}_{\Delta x_{N}, \Delta z_{N}} \mathrm{Cov}\left[q\left(x,z,\tau_N \right),q\left(x+\Delta x,z+\Delta z,\tau_N +\Delta \tau \right)\right]$.  The three marked velocities are $c=\{0.59, 0.58, 0.54\}$, or equivalently $c/u_\tau = \{9.2, 9.0, 8.4\}$.}
            \label{fig:c_Omega}
        \end{figure}

To understand the last observation, we note from the formula \eqref{eq:bdrydefnm} that a persistent correlation between the two fields, $\Omega_i^z$ and $\sigma_i$, is required to produce a substantial contribution to the space-time integral for $\mathcal{B}_{Ni}^z$. 
As observed already by \cite{Lighthill}
\begin{equation} 
    \sigma_z=\partial p/\partial x, 
    \qquad 
    \sigma_x = -\partial p/\partial z, 
\end{equation}
whereas $\sigma_y=-\nu\partial\omega_y/\partial y$ is not directly related to pressure. 
Furthermore, it was discovered by \cite{kim1993propagation} from numerical simulations that pressure propagates in the buffer and viscous layers with a phase speed of about 13$\,u_\tau$, which is higher than the phase speed of the velocity perturbations in this region, which is only 9.5$\,u_\tau$.  Thus, if the adjoint fields $\Omega_i^z$ for $i=x,y,z$ 
all propagate with the phase speed of the velocity perturbations, then they will become rapidly uncorrelated with $\sigma_z'$ and $\sigma_x'$, but may remain correlated with $\sigma_y'$ over long times.  Because the fluctuating source terms $\sigma_i'$ have zero means with equal contributions of both signs, the decorrelation will lead to strongly depleted contributions to $\mathcal{B}_{Nz'}^z$ and $\mathcal{B}_{Nx}^z$ compared with $\mathcal{B}_{Ny}^z$ at large backward times. 
To test this hypothesis, we measured the phase speeds of all of the wall fields, vorticity source $\sigma_i(\boldsymbol{x},t)$ and wall adjoint vorticity $\Omega_{Ni}^z(\boldsymbol{x},t)$ for $i=x,y,z$. 
For the source, since the field is homogeneous in the streamwise direction and statistically stationary in time, we used a standard method of space-time correlations.  The results are plotted in figure~\ref{fig:c_flux}, and show that, as expected, $\sigma_x'$ and $\sigma_z'$ have phase speeds close to $13\,u_\tau$ whereas $\sigma_y'$ has phase speed about $9\,u_\tau$.  Since the fields of $\Omega_i^z$ are not streamwise homogeneous or statistically stationary, we used a different method.  
We evaluated the horizontal shifts that maximize the correlation between fields at successive reverse times, and thus evaluated the upstream displacement of the adjoint field as a function of $\tau$, and then averaged this quantity over the ensemble of 109 independent samples of local stress maxima.  The results are plotted in figure~\ref{fig:c_Omega}, and show that the phase speeds of the adjoint are $\sim 9.5\,u_\tau$, similar to those of the near-wall velocity. 
Figures \ref{fig:c_flux} and \ref{fig:c_Omega}, together, support our proposed explanation why $\mathcal{B}_{Ny}^z$ is the dominant part of the wall vorticity source contribution to the target vorticity.  

The present analysis demonstrates that the stress extrema in channel flow are primarily due to stretching of interior near-wall vorticity, for reverse times on the order of $\tau^+ \simeq 347$.  The contribution of the vorticity flux at the wall, or the Lighthill source term, is relatively small and does not necessarily have the correct sign for individual events.  On average, this term is small, with half of its mean arising from the externally applied streamwise pressure gradient.  More interestingly, the distribution of the boundary term is essentially set by the flux of wall-normal vorticity that is subsequently tilted to generate a spanwise stress.  The analysis highlights the incisive utility of the adjoint vorticity equation \eqref{eq:operator}, and its value in tracing back the origin of vorticity in viscous incompressible flows in a manner as insightful as the original inviscid formulation but with a rigorous account for the influence of viscosity.


\section{Conclusion}
\label{sec:conclusion}

Vorticity and vortex dynamics are central to fluid dynamics as testified by modern monographs devoted  to their study \citep{saffman1995vortex, Majda_Bertozzi_2001, wu2007vorticity}. 
In inviscid flows, a vortex line is transported as a material line. Its advection, tilting, and stretching are represented exactly by the deformation matrix.  Vorticity being an invariant of both the forward or back in time flow maps, we can track it forward in time to examine its evolution or backward in time to identify its origin.  Once viscosity is present, however, this elegant description is no longer possible.  A vortex line is no longer a material line.  Instead, the vorticity of a point influences all other material points in forward time, and is dependent on all other points in the field at earlier time due to viscous diffusion.   

The forward evolution of vorticity in a viscous fluid is governed by the well-established Helmholtz equation. Much more recently, a stochastic Lagrangian approach was introduced to trace back the origin of vorticity \citep{ConstantinIyer08,Constantin2011}.  In that approach, stochastic particles are released from the points of interest, and are tracked backward in time. The vorticity at the point and time of release of the particles can then be evaluated as the expectation of the vorticities of these particles at any earlier time, each stretched and tilted according to the total deformation that is accumulated over the associated stochastic trajectory. Since this formula is valid for any earlier time, the terminal vorticity is considered an invariant of the stochastic backward trajectories \citep{Eyink2020_theory}. This approach requires specialized algorithms and has a high computational cost, since the number of necessary stochastic particles grows with the backward time horizon of interest and the Reynolds number. 

In this work, we introduced an Eulerian adjoint method to determine the backward-in-time origin of vorticity in viscous incompressible flows.  A key assumption was to freeze the velocity field along which the adjoint is evolved back in time, and the particular adjoint representation that we present is distinguished by being mathematically equivalent to the stochastic Lagrangian representation.  This equivalence underlies the physical interpretation of the adjoint as a volume density of mean Lagrangian deformation.   Forward-adjoint duality relates the terminal vorticity to the stretching, tilting, and boundary contributions earlier in time. 
We considered both Dirichlet and Neumann boundary conditions at solid boundaries.  In the former case, the adjoint samples the forward wall vorticity itself, and does not account for its origin further back in time. In the Neumann case, the adjoint samples the boundary vorticity flux.  While both approaches can be used to analyze the back-in-time origin of internal vorticity within the fluid, only the Neumann case can trace back the origin of vorticity directly on a wall, or the wall stress. 

In order to demonstrate the utility of the formulation, we applied it to examine high-stress events in turbulent channel flow, at $Re_\tau=180$. To compare the Dirichlet and Neumann cases, we analyzed the vorticity at $y^+=5$ above the local wall-shear-stress maximum, for an ensemble of 109 independent events. In the Dirichlet case, the majority of the vorticity can be attributed to the wall, arising within a time of about fifty to one hundred viscous units.  In contrast, the Neumann interpretation shows that the vorticity at this height is due to stretching of internal spanwise vorticity, and that the wall vorticity flux makes a relatively small contribution, which can be of the opposite sign.  These results accord well with the conjectures of \cite{Lighthill} on the origin of strong near-wall vorticity in turbulent wall-bounded flows.

The high-stress events directly on the wall were also studied, using the Neumann formulation.  The analysis examined the origin of the spanwise vorticity on the wall, and identified its origin in backward time.  Even at the wall, the peaks in the spanwise vorticity are on average  $75\%$ due to stretching of earlier in time interior vorticity, as far back as $\tau^+ = 347$.  In comparison, the vorticity flux over this entire duration accounts for only $25\%$ of the terminal value, on average. We show that about half of this small percentage is due to the externally applied streamwise pressure gradient. The distribution of the boundary term is, however, set by the flux of wall-normal vorticity, which is subsequently tilted.  The other two components of the fluctuating vorticity flux, which are associated with the fluctuating pressure gradient on the wall, make a relatively small contribution due to their fast phase speed relative to the adjoint field.  These components can, however, be expected to play a role when significant pressure gradient effects are at play, for example in bluff-body flows.

\appendix 


	\section{Equivalence with stochastic Lagrangian approach}\label{math}

    \subsection{Neumann boundary conditions}
	\label{sec:stochasticN}

We here briefly review the stochastic Lagrangian representation of vorticity with Neumann boundary condition which was 
introduced by \cite{wang_eyink_zaki_2022} and then establish its equivalence to the adjoint formulation. The 
representation of \cite{wang_eyink_zaki_2022} employed the backward stochastic Lagrangian flow with reflection at the boundary, 
which satisfies the backward It${\bar{\rm o}}$ SDE (with $dt<0$):
\begin{equation}
\hat{d}\widetilde{{\boldsymbol A}}_T^t({\boldsymbol x}_f)={\boldsymbol u}(\widetilde{{\boldsymbol A}}_T^t({\boldsymbol x}_f),t)dt 
+\sqrt{2\nu}\,\hat{d}\widetilde{{\boldsymbol W}}(t)+\nu {\boldsymbol n}(\widetilde{{\boldsymbol A}}_T^t({\boldsymbol x}_f))\hat{d}\tilde{\ell}^t_T({\boldsymbol x}_f).
\label{back_eq_refl} \end{equation}
Here $\boldsymbol{n}$ is the outward-pointing normal as in the main text and 
$\tilde{\ell}^t_T({\boldsymbol x}_f)$ is the backward-in-time boundary local time 
density which implements the inward reflection and which is defined formally as
\begin{equation}
\tilde{\ell}^t_T({\boldsymbol x}_f)
=\int_T^t dr\, \int_{\partial D} dS({\boldsymbol x}) \, \delta^3({\boldsymbol x}-\widetilde{{\boldsymbol A}}_T^r({\boldsymbol x}_f)), 
\quad s<t<T.
\label{loc_time_back} \end{equation}
We follow here the conventions of \cite{drivas2017lagrangianII}, who defined the boundary local time density 
to be negative and backward-in-time non-increasing. In fact, this quantity only decreases (backward in time) 
when the particle hits the boundary. The backward It${\bar{\rm o}}$ equation \eqref{back_eq_refl} with reflection
at the boundary can be transformed to a forward It${\bar{\rm o}}$ equation by 
time-reversal \citep{cattiaux1988time,petit1997time}. The reversed backward flow 
$\widetilde{\underline{{\boldsymbol A}}}_s^{\underline{t}}({\boldsymbol x}_f)$ defined by $\widetilde{\underline{{\boldsymbol A}}}_s^{\underline{t}}({\boldsymbol x}_f):=\widetilde{{\boldsymbol A}}_T^t({\boldsymbol x}_f)$ with 
${\underline t}=T+s-t,$ satisfies a forward It${\bar{\rm o}}$ SDE 
\begin{equation}
d\widetilde{\underline{{\boldsymbol A}}}_s^{\underline t}({\boldsymbol x}_f)=\underline{{\boldsymbol u}}
(\widetilde{{\underline{\boldsymbol A}}}_s^{\underline t}({\boldsymbol x}_f),{\underline t})
d{\underline t} 
+\sqrt{2\nu}\,d\widetilde{\underline{{\boldsymbol W}}}({\underline t})
-\nu {\boldsymbol n}(\widetilde{\underline{{\boldsymbol A}}}_s^{\underline t}({\boldsymbol x}_f))d\tilde{\underline{\ell}}_s^{\underline t}({\boldsymbol x}_f) 
\label{rev_back_eq_ref}
\end{equation} 
with time-reversed velocity field $
 \underline{{\boldsymbol u}}({\boldsymbol x},{\underline t)}:=
 -{\boldsymbol u}({\boldsymbol x},t),$ time-reversed Brownian motion 
$\widetilde{\underline{{\boldsymbol W}}}({\underline t}):=-\widetilde{\underline{{\boldsymbol W}}}(t)$ and boundary local time density, non-negative and non-decreasing:  
$$ \tilde{\underline{\ell}}_s^{\underline t}({\boldsymbol x}_f)
=\int_s^{\underline t} dr\, \int_{\partial D} dS({\boldsymbol y}) \, \delta^3({\boldsymbol y}-\widetilde{\underline{{\boldsymbol A}}}_s^r({\boldsymbol x}_f)), 
\quad s<{\underline t}<T$$
so that 
$
\widetilde{{\boldsymbol \ell}}_T^t({\boldsymbol x}_f)=\widetilde{\underline{{\boldsymbol \ell}}}_s^{t}({\boldsymbol x}_f)
-\widetilde{\underline{{\boldsymbol \ell}}}_s^T({\boldsymbol x}_f).$
This differs from the natural forward stochastic flow $\widetilde{{\boldsymbol X}}_s^t({\boldsymbol a})$
reflected at the boundary, which is inverted by the backward flow, $\widetilde{{\boldsymbol X}}_s^t\circ \widetilde{{\boldsymbol A}}_t^s={\rm Id},$ and which satisfies instead the forward It${\bar{\rm o}}$ equation 
$$ d\widetilde{{\boldsymbol X}}_s^t({\boldsymbol a})={\boldsymbol u}(\widetilde{{\boldsymbol X}}_s^t({\boldsymbol a}),t)dt 
+\sqrt{2\nu}\,d\widetilde{{\boldsymbol W}}(t)-\nu {\boldsymbol n}(\widetilde{{\boldsymbol X}}_s^t({\boldsymbol a}))
\hat{d}\tilde{\ell}_s^t({\boldsymbol a}) $$ 
with its boundary local time density 
$$ \tilde{\ell}_s^t({\boldsymbol a})
=\int_s^t dr\, \int_{\partial D} dS({\boldsymbol y}) \, \delta^3({\boldsymbol y}-\widetilde{{\boldsymbol X}}_s^r({\boldsymbol a})), 
\quad s<t<T$$
The transition probabilities of $\widetilde{\underline{{\boldsymbol A}}}_s^{\underline t}$ and $\widetilde{{\boldsymbol X}}_s^t$ 
are related, however, by
\begin{equation} 
\underline{p}({\boldsymbol x}',{\underline t}'|{\boldsymbol x},{\underline t}) = p({\boldsymbol x},t|{\boldsymbol x}',t'), 
\label{trans-pdf-rev} \end{equation} 
as seen by comparing the forward Kolmogorov equation for the process
$\widetilde{\underline{{\boldsymbol A}}}_s^{\underline t}({\boldsymbol x}_f)$ 
with the backward Kolmogorov equation for the process $\widetilde{{\boldsymbol X}}_s^t({\boldsymbol x}).$ Thus  
$$ \int_D \underline{p}({\boldsymbol x}',{\underline t}'|{\boldsymbol x},{\underline t}) \, d^3x
= \int_D p({\boldsymbol x},t|{\boldsymbol x}',t')\, d^3x' = 1. $$
These relations imply therefore the important fact that both the forward and the backward stochastic flows 
with reflection at the boundary preserve the fluid volume. 

The formula of \cite{wang_eyink_zaki_2022} represents vorticity in terms 
of the backward stochastic Lagrangian flow and its boundary local time density as 
\begin{equation}
{\boldsymbol \omega}({\boldsymbol x}_f,T)={\mathbb E}\left[\widetilde{{\boldsymbol D}}_T^s({\boldsymbol x}_f)  
{\boldsymbol \omega}(\widetilde{{\boldsymbol A}}_T^s({\boldsymbol x}_f),s) + 
\int_s^T \widetilde{{\boldsymbol D}}_T^r({\boldsymbol x}_f) 
{\boldsymbol \sigma}(\widetilde{{\boldsymbol A}}_T^r({\boldsymbol x}_f),r)
\hat{d}\tilde{\ell}^r_T({\boldsymbol x}_f) \right]
\label{omega_rep_refl} \end{equation} 
where ${\boldsymbol \sigma}=\nu \boldsymbol{n} \cdot \boldsymbol{\nabla} \boldsymbol{\omega}$
is the boundary vorticity source of \cite{Lighthill,panton1984incompressible} and 
$\widetilde{{\boldsymbol D}}_T^s({\boldsymbol x}_f)$ is the ``deformation matrix''
which satisfies the ODE \eqref{Def_eq} backward in time. It is an interesting question whether the martingale identified by \cite{wang_eyink_zaki_2022} has geometric meaning, or, more precisely, whether one may formulate their representation using the geometric deformation matrix defined by $\widetilde{{\boldsymbol D}}_t^s({\boldsymbol x}_f):=\left.
({\boldsymbol \nabla_a}\widetilde{{\boldsymbol X}}^t_s({\boldsymbol a}))^\top 
\right|_{{\boldsymbol a}=\widetilde{{\boldsymbol A}}_t^s({\boldsymbol x}_f)}.$ It has been 
proved by \cite{andres2011pathwise} that this derivative matrix exists pathwise, but it 
satisfies the matrix ODE \eqref{Def_eq} only between collisions and it is 
projected to have no normal component after a collision, i.e. 
${\boldsymbol \nabla_a}\widetilde{{\boldsymbol X}}^{t+}_s({\boldsymbol a})=
{\boldsymbol \nabla_a}\widetilde{{\boldsymbol X}}^{t-}_s({\boldsymbol a})
({\boldsymbol I}-{\boldsymbol n}{\boldsymbol n})$ 
for $t$ any collision time where $d\widetilde{\ell}^t_s({\boldsymbol a})>0.$ 
The reason for this projection is that the particle reaches an extremum of position in the wall-normal direction at the instant of reflection. 
Here we follow \cite{wang_eyink_zaki_2022} in 
defining $\widetilde{{\boldsymbol D}}_t^s({\boldsymbol x}_f)$ as the solution of 
\eqref{Def_eq} with $\widetilde{{\boldsymbol D}}_t^t({\boldsymbol x}_f)
={\boldsymbol I},$ which then represents cumulative stretching between
reflections along the trajectory
and which makes the quantity inside the expectation in \eqref{omega_rep_refl} a backward martingale. 

With the definitions 
of ${\boldsymbol \Upomega}$ in eq.\eqref{Omega_def} and $\tilde{\ell}^t_T({\boldsymbol x}_f)$
in eq.\eqref{loc_time_back}, we can rewrite the stochastic Lagrangian representation 
\eqref{omega_rep_refl} as 
$$ {\boldsymbol \omega}({\boldsymbol x}_f,T)
= \int_D d^3x \, {\boldsymbol \Upomega}({\boldsymbol x},s) 
{\boldsymbol \omega}({\boldsymbol x},s) 
+\int_s^T dr \int_{\partial D} dS({\boldsymbol y}) \, 
{\boldsymbol \Upomega}({\boldsymbol y},r) 
{\boldsymbol \sigma}({\boldsymbol y},r).  
$$
This agrees with the adjoint formulation. Furthermore, ${\boldsymbol \Omega}$ satisfies the adjoint vorticity 
equation in the flow interior by the same proof as before and we must only establish its boundary condition. 
Here it is helpful rewrite the stochastic representation of the adjoint vorticity field as
$$ {\boldsymbol \Upomega}({\boldsymbol x},t) := {\mathbb E}\left[ \widetilde{{\boldsymbol D}}_T^{t}({\boldsymbol x}_f)
\delta^3({\boldsymbol x}_f-\widetilde{{\boldsymbol X}}^T_t({\boldsymbol x}))\right]$$
in terms of the inverse map $\widetilde{{\boldsymbol X}}^T_t,$
using the fact that the stochastic flow for an incompressible velocity field preserves volume and 
that ${\rm det}[{\boldsymbol\nabla} \widetilde{{\boldsymbol X}}^T_t({\boldsymbol x}))]=1$
for a.e. ${\boldsymbol x}\in D$ with probability one. In that case we obtain by the chain rule 
$$ \frac{\partial}{\partial n} {\boldsymbol \Upomega}({\boldsymbol y},t)=
-{\mathbb E}\left[ \widetilde{{\boldsymbol D}}_T^{t}({\boldsymbol x}_f) 
\left(\frac{\partial}{\partial n}\widetilde{{\boldsymbol X}}^T_t({\boldsymbol y}){\boldsymbol\cdot\nabla}_{{\boldsymbol x}_f}
\right)
\delta^3({\boldsymbol x}_f-\widetilde{{\boldsymbol X}}^T_t({\boldsymbol y}))\right]={\boldsymbol 0},\quad 
{\boldsymbol y}\in \partial D,$$
since $\frac{\partial}{\partial n}\widetilde{{\boldsymbol X}}^T_t({\boldsymbol y})={\boldsymbol 0}$ almost surely. 
See \cite{andres2011pathwise}, \S 3.4.

    \subsection{Dirichlet boundary conditions}
	\label{sec:stochasticD}

We next review the stochastic Lagrangian representation of vorticity with Dirichlet b.c. which was 
introduced by \cite{Constantin2011} and establish its equivalence to the adjoint formulation. The 
representation of \cite{Constantin2011} employed the backward stochastic Lagrangian flow which 
is stopped at the boundary. The flow map for this process satisfies the backward It${\bar{\rm o}}$ SDE 
\eqref{back_eq} that was introduced earlier, but now the evolution is stopped at the first hitting time 
(backward in time):  
$$ \widetilde{\sigma}({\boldsymbol x}_f,T)=\sup \{t<T:\, 
\widetilde{{\boldsymbol A}}_T^t({\boldsymbol x}_f)\in \partial V\}. $$
The representation of \cite{Constantin2011} then takes the form
\begin{equation}
{\boldsymbol \omega}({\boldsymbol x}_f,T)={\mathbb E}\left[
\widetilde{{\boldsymbol D}}_T^{\widetilde{\sigma}({\boldsymbol x}_f,T)\vee s}({\boldsymbol x}_f)  
{\boldsymbol \omega}(\widetilde{{\boldsymbol A}}_T^{\widetilde{\sigma}({\boldsymbol x}_f,T)\vee s}({\boldsymbol x}_f),
\widetilde{\sigma}({\boldsymbol x}_f,T)\vee s)\right]. 
\label{omega_rep_stop} \end{equation} 
We now show that this representation coincides with the adjoint formulation. 

To begin, we must discuss some geometry. It is useful to assume that the domain $D$ is specified as the set $\{\varphi <0\}$ 
for a smooth function $\varphi$ with $\partial D$ given as the level set $\{\varphi=0\}.$ 
A simple example would be $D$ a ball of radius $R$ centered at the origin and 
$\varphi(x,y,z)=\log(\sqrt{x^2+y^2+z^2}/R).$ Useful coordinates in the domain $D$ 
are then chosen as the values $\psi$ of $\varphi$ and arbitrary smooth coordinates ${\boldsymbol y}={\boldsymbol \Sigma}({\boldsymbol x})$ 
on the level sets $\{\varphi=\psi\}.$ Note that the normal unit vector on level sets is 
${\boldsymbol n}={\boldsymbol \nabla}\varphi/|{\boldsymbol \nabla}\varphi|$ and the normal distance between infinitesimally separated levels is 
$dn=d\psi/|{\boldsymbol \nabla}\varphi|$ so that $d^3x=dn\,dS({\boldsymbol y})=d\psi\, dS({\boldsymbol y})/|{\boldsymbol \nabla}\varphi|.$ 
Locally, we may always solve the equation $\varphi(x,y,z)=\psi$ to find one of the coordinates as a function of the other two, e.g. if $\varphi_z\neq 0,$ then $z=\zeta(x,y;\psi)$ and in that case ${\boldsymbol y}=(x,y)={\boldsymbol \Sigma}(x,y,z).$ 
It then follows that $d{\boldsymbol S}=(\hat{{\boldsymbol x}} + \zeta_x \hat{{\boldsymbol z}})dx
{\boldsymbol \times} (\hat{{\boldsymbol y}} + \zeta_y \hat{{\boldsymbol z}})dy=\frac{{\boldsymbol \nabla}\varphi}{\varphi_z}dx\,dy
={\boldsymbol n} \frac{|{\boldsymbol \nabla}\varphi|}{|\varphi_z|}dx\,dy = {\boldsymbol n}\, dS({\boldsymbol y}).$

With these preliminaries, we introduce into the stochastic representation \eqref{omega_rep_stop}
the sum of the two characteristic functions
$$ 1_{\{{\rm dist}(\widetilde{{\boldsymbol A}}_T^s({\boldsymbol x}_f)),\partial D)>0\}}
= \int_D d^3x \ \delta^3({\boldsymbol x} - \widetilde{{\boldsymbol A}}_T^s({\boldsymbol x}_f)), $$
and 
$$ 1_{\{{\rm dist}(\widetilde{{\boldsymbol A}}_T^s({\boldsymbol x}_f)),\partial D)=0\}}
= \int_s^T dr \int_{\partial D} dS({\boldsymbol y})\ \delta^2_S({\boldsymbol y},{\boldsymbol \Sigma}(\widetilde{{\boldsymbol A}}_T^{r}({\boldsymbol x}_f)))
\delta(r-\widetilde{\sigma}({\boldsymbol x}_f,T)), $$
where $\delta^2_S({\boldsymbol y},{\boldsymbol y}')$ is surface delta function given e.g. by $\delta^2_S({\boldsymbol y},{\boldsymbol y}')=\frac{|\varphi_z|}{|{\boldsymbol \nabla}\varphi|}\delta(x-x')\delta(y-y').$
One then obtains the result 
$$ {\boldsymbol \omega}({\boldsymbol x}_f,T)
= \int_D  d^3x \,  {\boldsymbol \Upomega}({\boldsymbol x},s) 
{\boldsymbol \omega}({\boldsymbol x},s)
+\int_s^T dr \int_{\partial D} dS({\boldsymbol y}) \, 
{\boldsymbol \Upomega}^{\prime}({\boldsymbol y},r) 
{\boldsymbol \omega}({\boldsymbol y},r) 
$$
where ${\boldsymbol \Upomega}$ is defined in \eqref{Omega_def} and we have defined also 
\begin{equation}
{\boldsymbol \Upomega}'({\boldsymbol y},r) := {\mathbb E}\left[ \widetilde{{\boldsymbol D}}_T^{r}({\boldsymbol x}_f)
\delta^2_S({\boldsymbol y},{\boldsymbol \Sigma}(\widetilde{{\boldsymbol A}}_T^{r}({\boldsymbol x}_f)))
\delta(r-\widetilde{\sigma}({\boldsymbol x}_f,T))
\right]
\label{Omega_pr_def} \end{equation} 
Note that ${\boldsymbol \Upomega}={\boldsymbol 0}$ on $\partial D,$ since the probability density 
for the stopped/absorbed/killed process vanishes there. For example, see 
\cite{sobczyk1990stochastic}, \S 6.2.2. 

To conclude, we must show that ${\boldsymbol \Upomega}'({\boldsymbol y},r) 
= -\nu \frac{\partial}{\partial n} {\boldsymbol \Upomega}({\boldsymbol y},r).$
We first calculate $ \frac{\partial}{\partial n}{\boldsymbol \Upomega}({\boldsymbol y},r).$
Noting that 
$$ \delta^3({\boldsymbol x}-\widetilde{{\boldsymbol A}}_T^{r}({\boldsymbol x}_f)) 
= \delta^2_S({\boldsymbol y},{\boldsymbol \Sigma}(\widetilde{{\boldsymbol A}}_T^{r}({\boldsymbol x}_f)))
\cdot |{\boldsymbol \nabla}\varphi|\delta(\psi-\varphi(\widetilde{{\boldsymbol A}}_T^{r}({\boldsymbol x}_f)),$$
we see from eq.\eqref{Omega_def} that 
${\boldsymbol \Upomega}({\boldsymbol x},r)={\boldsymbol \Upomega}({\boldsymbol y},\psi,r)$ is given by 
$$ {\boldsymbol \Upomega}({\boldsymbol y},\psi,r) = {\mathbb E}\left[ \widetilde{{\boldsymbol D}}_T^{r}({\boldsymbol x}_f)\,
|{\boldsymbol \nabla}\varphi(\widetilde{{\boldsymbol A}}_T^{r}({\boldsymbol x}_f))|\,
\delta^2_S({\boldsymbol y},{\boldsymbol \Sigma}(\widetilde{{\boldsymbol A}}_T^{r}({\boldsymbol x}_f)))
\delta(\psi-\varphi(\widetilde{{\boldsymbol A}}_T^{r}({\boldsymbol x}_f))
\right].$$
Using then $\frac{\partial}{\partial n}=|{\boldsymbol \nabla}\varphi|\frac{\partial}{\partial \psi},$ we see that at the boundary 
$\partial D$ where $\psi=0$
$$ \frac{\partial}{\partial n}{\boldsymbol \Upomega}({\boldsymbol y},r) = -
{\mathbb E}\left[ \widetilde{{\boldsymbol D}}_T^{r}({\boldsymbol x}_f)\,
|{\boldsymbol \nabla}\varphi(\widetilde{{\boldsymbol A}}_T^{r}({\boldsymbol x}_f))|^2\,
\delta^2_S({\boldsymbol y},{\boldsymbol \Sigma}(\widetilde{{\boldsymbol A}}_T^{r}({\boldsymbol x}_f)))
\delta'(\varphi(\widetilde{{\boldsymbol A}}_T^{r}({\boldsymbol x}_f))
\right].$$
We next calculate ${\boldsymbol \Upomega}'({\boldsymbol y},r).$  For this purpose we must use the following basic relation 
$$ 1_{\{\widetilde{\sigma}({\boldsymbol x}_f,T)<r\}} = 1_{\{{\rm dist}(\widetilde{{\boldsymbol A}}_T^r({\boldsymbol x}_f)),\partial D)>0\}}. 
$$
Thus, we find by differentiation in $r$ that 
$$
\delta(r-\widetilde{\sigma}({\boldsymbol x}_f,T)) =
\frac{d}{dr} 1_{\{{\rm dist}(\widetilde{{\boldsymbol A}}_T^r({\boldsymbol x}_f)),\partial D)>0\}} . 
$$  
We can use also
$$
1_{\{{\rm dist}(\widetilde{{\boldsymbol A}}_T^r({\boldsymbol x}_f)),\partial D)>0\}}
=\int^0_{-\infty} d\psi\, \delta (\psi-\varphi(\widetilde{{\boldsymbol A}}_T^r({\boldsymbol x}_f))). 
$$
Inserting these results into the definition \eqref{Omega_pr_def} of ${\boldsymbol \Upomega}'({\boldsymbol y},r),$ we find that  
\begin{equation} 
{\boldsymbol \Upomega}'({\boldsymbol y},r)dr= \int^0_{-\infty} d\psi\, 
{\mathbb E}\left[ \widetilde{{\boldsymbol D}}_T^{r}({\boldsymbol x}_f)
\delta^2_S({\boldsymbol y},{\boldsymbol \Sigma}(\widetilde{{\boldsymbol A}}_T^{r}({\boldsymbol x}_f)))
\hat{d}_r \delta (\psi-\varphi(\widetilde{{\boldsymbol A}}_T^r({\boldsymbol x}_f)))
\right].
\end{equation} 
We now use the backward It${\bar{\rm o}}$ rule to calculate 
\begin{eqnarray*}
\hat{d}_r \delta (\psi-\varphi(\widetilde{{\boldsymbol A}}_T^r)) & = & 
(\hat{d}_r\widetilde{{\boldsymbol A}}_T^r{\boldsymbol \cdot}{\boldsymbol \nabla}_{a}) 
\delta (\psi-\varphi(\widetilde{{\boldsymbol A}}_T^r)) -\nu\nabla^2_{\boldsymbol{a}} \delta (\psi-\varphi(\widetilde{{\boldsymbol A}}_T^r)) dr \cr
&=& -\delta'(\psi-\varphi(\widetilde{{\boldsymbol A}}_T^r)) \hat{d}_r\varphi(\widetilde{{\boldsymbol A}}_T^r)
-\nu \delta''(\psi-\varphi(\widetilde{{\boldsymbol A}}_T^r)) |{\boldsymbol \nabla}\varphi(\widetilde{{\boldsymbol A}}_T^{r})|^2 dr 
\end{eqnarray*} 
Substituting back into the formula for ${\boldsymbol \Upomega}'({\boldsymbol y},r)$ and performing the integral over $\psi$ yields
\begin{eqnarray*}
{\boldsymbol \Upomega}'({\boldsymbol y},r)dr &=&
-{\mathbb E}\left[ \widetilde{{\boldsymbol D}}_T^{r}
\delta^2_S({\boldsymbol y},{\boldsymbol \Sigma}(\widetilde{{\boldsymbol A}}_T^{r}))
\delta (\varphi(\widetilde{{\boldsymbol A}}_T^r))  \hat{d}_r\varphi(\widetilde{{\boldsymbol A}}_T^r)
\right] \cr 
&& + \nu {\mathbb E}\left[ \widetilde{{\boldsymbol D}}_T^{r}\,
|{\boldsymbol \nabla}\varphi(\widetilde{{\boldsymbol A}}_T^{r})|^2\,
\delta^2_S({\boldsymbol y},{\boldsymbol \Sigma}(\widetilde{{\boldsymbol A}}_T^{r}))
\delta'(\varphi(\widetilde{{\boldsymbol A}}_T^{r}))
\right] dr.
\end{eqnarray*}
However, the first term on the right vanishes because $\hat{d}_r\varphi(\widetilde{{\boldsymbol A}}_T^r)=0$ on the boundary $\partial D$
for the stopped process. Comparing with the previous expression for $\frac{\partial}{\partial n}{\boldsymbol \Upomega}({\boldsymbol y},r),$
we see that 
\begin{equation} {\boldsymbol \Upomega}'({\boldsymbol y},r) 
= -\nu \frac{\partial}{\partial n} {\boldsymbol \Upomega}({\boldsymbol y},r),
\label{Omega_pr_eq}
\end{equation}
exactly as required. 
Equations \eqref{Omega_pr_def} and \eqref{Omega_pr_eq} thus demonstrate that the adjoint boundary term $-\nu\partial \boldsymbol{\Upomega}/\partial n$ in \eqref{eq:operator} is interpretable as the density of mean deformation for stochastic Lagrangian particles hitting the wall per unit area and per unit time.

{\section{Statistical contributions to target vorticity at different backward times for the ensemble of high stress events}\label{statcontr}}

\begin{figure}
    \centering
        \includegraphics[width=0.7\textwidth]{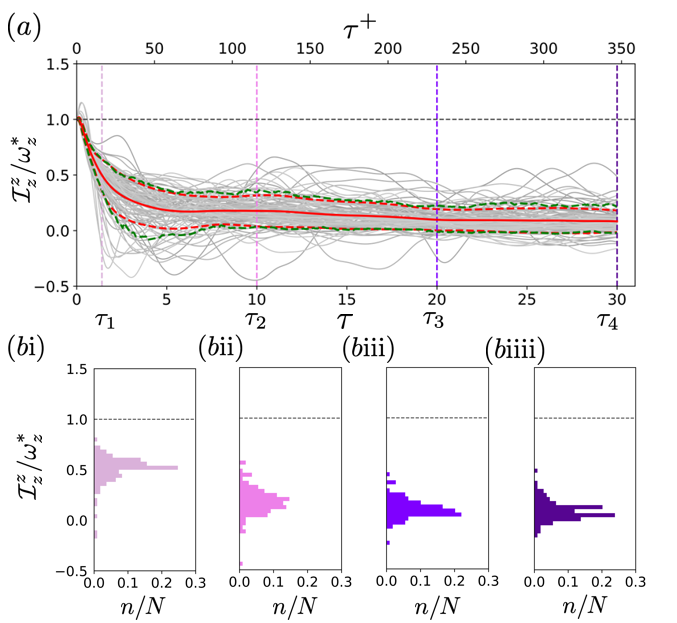}
        \caption{
        Ensemble of 109 target vorticity events, at $y^+=5$ above a wall-stress maximum, tracked back in time using the Dirichlet condition.  $(a)$ Gray lines are the interior vorticity stretching ${\mathcal I}^z_z=\Omega_{z}^{z} \omega_{z}$ normalized by the target values $\omega_z^* = \omega_z\left(\boldsymbol{x}_f, T\right)$. ({\color{red}$\lline$}) Ensemble-averaged value; 
        ({\color{red}$\dashed$}) $\pm$ the standard deviation; ({\color{green}$\dashed$}) 10\% and 90\% percentiles.  
        $(b)$ Histograms of the interior stretching contribution at $(\mathrm{i})$ $\tau_1=1.4$, $(\mathrm{ii})$ $\tau_2=10$, $(\mathrm{iii})$ $\tau_3=20$ and $(\mathrm{iiii})$ $\tau_4=30$.}
    \label{fig:EnsembleDirichletspanz}
    \vspace*{12pt}
    \centering
        \includegraphics[width=0.7\textwidth]{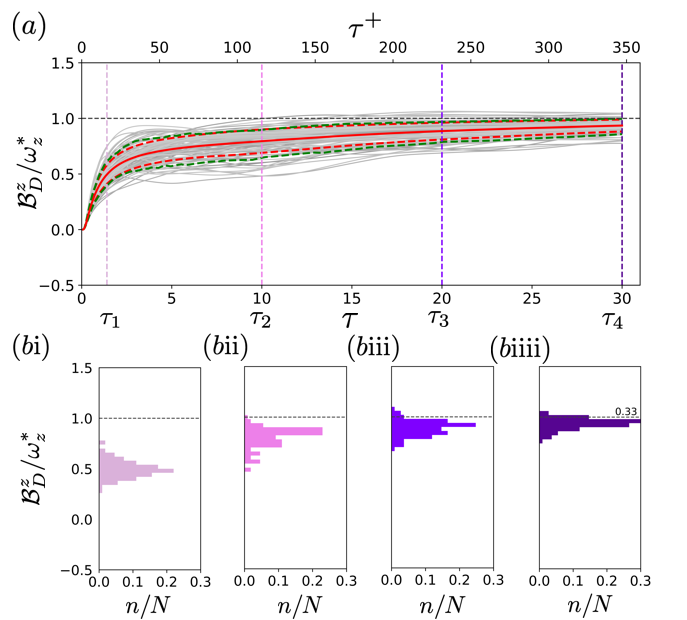}
        \caption{
        Same as figure \ref{fig:EnsembleDirichletspanz}, but here for the boundary 
        vorticity contribution ${\mathcal B}^z_D$ to the target vorticity with the Dirichlet condition.  }
    \label{fig:EnsembleDirichletwall}
\end{figure}

\begin{figure}
    \centering
        \includegraphics[width=0.7\textwidth]{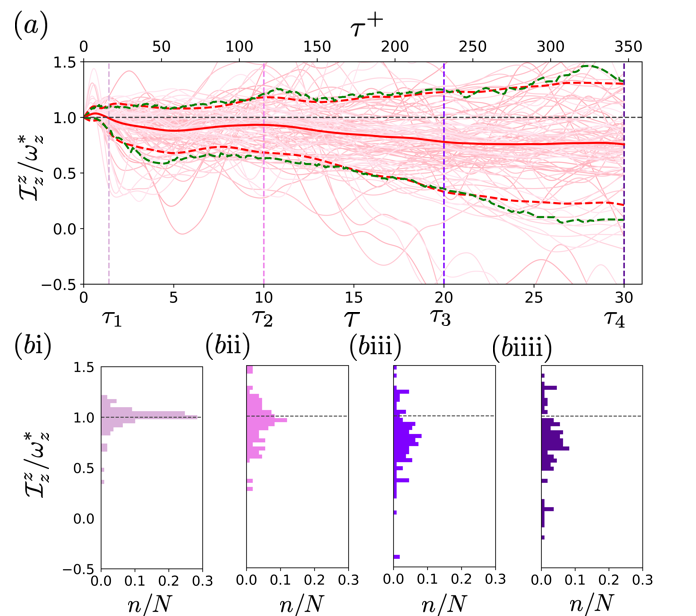}
        \caption{
        Ensemble of 109 target vorticity events, at $y^+=5$ above a wall-stress maximum, tracked back in time using the Neumann condition.  $(a)$ Pink lines are the interior vorticity stretching ${\mathcal I}^z_Z=\Omega_{z}^{z} \omega_{z}$ normalized by the target values $\omega_z^* = \omega_z\left(\boldsymbol{x}_f, T\right)$. ({\color{red}$\lline$}) Ensemble-averaged value; 
        ({\color{red}$\dashed$}) $\pm$ the standard deviation; ({\color{green}$\dashed$}) 10\% and 90\% percentiles.  
        $(b)$ Histograms of the interior stretching contribution at $(\mathrm{i})$ $\tau_1=1.4$, $(\mathrm{ii})$ $\tau_2=10$, $(\mathrm{iii})$ $\tau_3=20$ and $(\mathrm{iiii})$ $\tau_4=30$. }
    \label{fig:EnsembleNeumannspanz}
    \vspace*{12pt}
    \centering
        \includegraphics[width=0.7\textwidth]{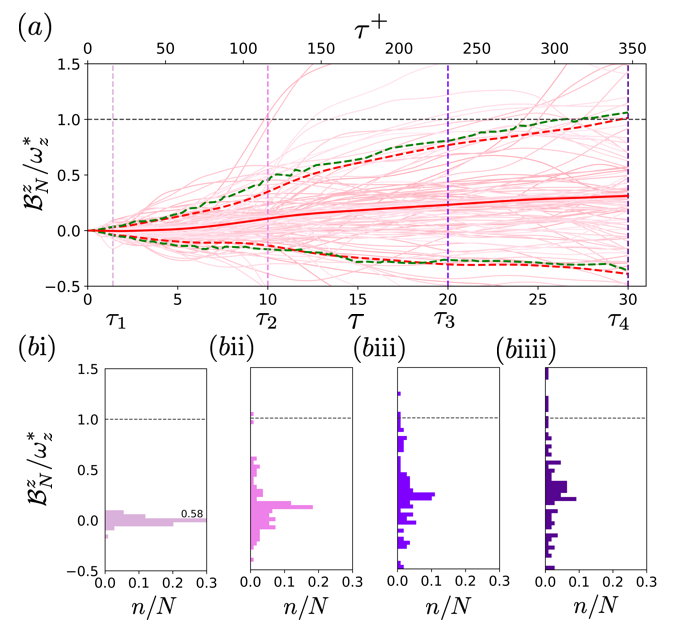}
        \caption{
        Same as figure \ref{fig:EnsembleNeumannspanz}, but here for the boundary 
        vorticity flux contribution ${\mathcal B}^z_N$ to the target vorticity with the Neumann condition. }
    \label{fig:EnsembleNeumannwall}
\end{figure}

In the main text we plotted as functions of backward time the two largest fractional contributions $\mathcal{I}^z_z$ and $\mathcal{B}_{D/N}^z$ to the target vorticity for the ensemble of 
high shear-stress events at $y^+=5,$ with Dirichlet boundary conditions in figure \ref{fig:EnsembleDirichlet} and with Neumann conditions in figure \ref{fig:EnsembleNeumann}. In this appendix we provide a more 
detailed view of the statistical contributions by plotting their probability distributions at selected backward times.

For Dirichlet boundary conditions, we plot the histories of the vorticity-stretching contributions in figure \ref{fig:EnsembleDirichletspanz}(a), identical to those in figure \ref{fig:EnsembleDirichlet}$(a)$ of the main text, and we plot the histories of the boundary contributions in figure \ref{fig:EnsembleDirichletwall}$(a)$, identical to those in figure \ref{fig:EnsembleDirichlet}$(b)$. 
In addition to the mean plus or minus the standard deviation, we also report the 10\% and 90\% percentiles using green dashed lines. The curves for the mean plus/minus standard deviation and for the percentiles are quite similar, both giving a good measure of the ensemble variation.  Most importantly, we plot in these new figures the histograms of the fractional vorticity contributions at four selected backward times, for $\mathcal{I}^z_z$ in figure \ref{fig:EnsembleDirichletspanz}$(b)$ and for $\mathcal{B}_{D}^z$ in figure \ref{fig:EnsembleDirichletwall}$(b)$.  
The mean fractional contribution from spanwise stretching $\mathcal{I}^z_z$ exhibits a clear decrease from unity to zero in backward time,  
while the mean fractional contribution from spanwise wall vorticity $\mathcal{B}_{D}^z$ shows a complementary increase from zero to unity, consistent with our observations in section \ref{sec:statistical}.
Meanwhile, the histograms of both contributions have relatively small variance, roughly constant in time.
It is thus clear that the reconstruction of the vorticity with Dirichlet boundary conditions has rather low 
variability, 
with most histories very similar to the mean history and with very little differentiation of individual events.

The corresponding results for Neumann boundary conditions are much more revealing of the 
distinctive dynamics in individual events. We plot the histories of the vorticity-stretching contributions in figure \ref{fig:EnsembleNeumannspanz}(a), identical to those in figure \ref{fig:EnsembleNeumann}$(a)$ of the main text, and we plot the histories of the boundary contributions in figure \ref{fig:EnsembleNeumannwall}$(a)$, identical to those in figure \ref{fig:EnsembleNeumann}$(b)$.  
Just as for the Dirichlet condition, the mean plus or minus the standard deviation and the 10\% and 90\% percentiles are quite close to each other, although  the standard deviation with Neumann conditions is of the same order as the mean value. 
We again plot in panel $(b)$ of both figures the histograms of the two contributions at the four selected backward times. In contrast to the Dirichlet case, the mean contributions for Neuman conditions do not change as rapidly in backward time, while the distributions become much broader especially for the boundary term.  
Although the mathematical theory and common 
wisdom state that the target vorticity must ultimately arise entirely from the wall vorticity source, the time scale for the boundary contribution to dominate is extremely long
(equal in fact to the time required for the stochastic particles to mix uniformly over the flow domain). The very large variances of the two contributions from interior stretching and from the boundary flux are reflective of the highly individual histories of the different high-stress events. 
This difference from the Dirichlet case can be understood with reference to the stochastic Lagrangian interpretation.  In the Dirichlet case, particles that reach the wall sample the vorticity at that instant and stop,  whereas in the Neumann case particles 
are reflected from the wall after sampling the vorticity flux, because even the wall vorticity has history. That history is only accessible using the Neumann condition, which alone enables the study of the back-in-time origin of the surface stress.

    \par\bigskip
    \noindent
    \textbf{Acknowledgements.} The authors are grateful to Profs.\,D.\,Barkley and P.\,Luchini for inquiring about possible connections between the stochastic Cauchy invariant and an adjoint representation, which helped to motivate this work.  
   
   \par\bigskip
   \noindent
   \textbf{Funding.} Tamer A. Zaki acknowledges financial support from the Office of Naval Research (Grants  N00014-20-1-2715 and  N00014-21-1-2375). Gregory L. Eyink acknowledges the Simons Foundation for support (Targeted Grant No. MPS-663054, and Collaboration Grant No. MPS-1151713). Computational resources were provided by the Advanced Research Computing at Hopkins (ARCH) core facility.

    \par\bigskip
    \noindent
    \textbf{Declaration of interests.} 
    The authors report no conflict of interest.
     
    \par\bigskip
    \noindent
    \textbf{Author ORCIDs.} \\
    Gregory L. Eyink, \url{https://orcid.org/0000-0002-8656-7512} \\
    Tamer A. Zaki, \url{https://orcid.org/0000-0002-1979-7748}

\bibliographystyle{jfm}
\bibliography{reference}
	
\end{document}